\documentclass[a4paper,12pt]{article}

\usepackage{jheppub}                 
\usepackage{multicol}
\normalsize 

\usepackage{paralist}
\usepackage{graphicx}
\usepackage{enumitem}
\usepackage{bm}
\usepackage{color}
\usepackage{amsmath}
\usepackage[utf8]{inputenc}
\usepackage{placeins}
\usepackage[normalem]{ulem}
\usepackage{multirow}
\usepackage{bbold}

\newcommand{\ps}{p\hspace{-0.44em}/\hspace{0.06em}}
\newcommand{\eq}[1]{Eq.~(\ref{#1})}

\newcommand{\Bbar}{\,\overline{\!B}}

\newcommand{\beq} {\begin{equation}}
\newcommand{\eeq} {\end{equation}}
\newcommand{\bea} {\begin{eqnarray}}
\newcommand{\eea} {\end{eqnarray}}
\newcommand{\ba} {\begin{eqnarray*}}
\newcommand{\ea} {\end{eqnarray*}}
\newcommand{\GeV} {\,\text{GeV}}
\newcommand{\TeV} {\,\text{TeV}}

\newcommand{\hc}{\mathrm{h.c.}}

\def\PD      {\ensuremath{\mathrm{D}}\xspace}        
\def\Dbar    {\kern 0.2em\overline{\kern -0.2em \PD}{}\xspace}
\def\D       {\ensuremath{\PD}\xspace}

\def\Dz      {\ensuremath{\D^0}\xspace}
\def\Dzb     {\ensuremath{\Dbar^0}\xspace}

\def\PK      {\ensuremath{K}\xspace} 
\def\Kbar  {\kern 0.2em\overline{\kern -0.2em \PK}{}\xspace}

\def\PB      {\ensuremath{B}\xspace}         
\def\Bbar    {\ensuremath{\kern 0.18em\overline{\kern -0.18em \PB}{}}\xspace}

\usepackage[utf8]{inputenc} 
\usepackage[T1]{fontenc} 
\usepackage{lmodern}
\usepackage[british]{babel}
\usepackage{graphicx}
\graphicspath{{figs/}} 

\usepackage{microtype} 
\usepackage{subcaption} 
\usepackage{xfrac} 
\makeatletter
\g@addto@macro\@floatboxreset\centering
\makeatother
\usepackage{pdflscape} 

\numberwithin{equation}{section} 
\allowdisplaybreaks 
\usepackage{bbm} 

\usepackage[capitalise]{cleveref} 
\creflabelformat{equation}{\textup{#2#1#3}} 

\usepackage{booktabs} 
\usepackage{tabularx} 
\usepackage{longtable} 
\usepackage{array} 
\newcolumntype{L}{>{$}l<{$}} 

\usepackage[italic]{hepnames} 

\usepackage{braket} 
\usepackage{slashed} 
\usepackage{siunitx} 
\DeclareSIUnit\fb{\femto\barn}
\DeclareSIUnit\invps{\ps^{-1}}
\usepackage{suffix} 

\newcommand{\BigO}[1]{\ensuremath{\mathcal{O}\left(#1\right)}}

\newcommand{\Lagrangian}{\ensuremath{\mathcal{L}}\xspace}

\newcommand{\V}[1]{\ensuremath{V_{#1}^{}}} 
\WithSuffix\newcommand\V*[1]{\ensuremath{V_{#1}^*}}

\newcommand{\CHqone}{C^{(1)}_{Hq}}
\newcommand{\CHqonebracket}{\left[\CHqone\right]}
\newcommand{\CHqthree}{C^{(3)}_{Hq}}
\newcommand{\CHqthreebracket}{\left[\CHqthree\right]}
\newcommand{\CHud}{C_{Hud}}
\newcommand{\CHudbracket}{\left[\CHud\right]}
\newcommand{\program}[1]{\texttt{#1}}
\newcommand{\flavio}{\program{flavio}\xspace}
\newcommand{\smelli}{\program{smelli}\xspace}
\newcommand{\wilson}{\program{wilson}\xspace}
\newcommand{\mmeft}{\program{matchmakereft}\xspace}
\newcommand{\QuantumNumbers}[3]{(\mathbf{#1}, \mathbf{#2}, #3)}
\newcommand{\Yukawa}[1]{Y^{#1}}
\newcommand{\Yukawadagger}[1]{Y^{#1, \dagger}}

\usepackage{xcolor}

\preprint{PSI-PR-22-37,~ZU-TH-61/22,~KEK-TH-2480}
\vspace{-10mm}
\title{
Global Fit of Modified Quark Couplings to EW Gauge Bosons and Vector-Like Quarks in Light of the Cabibbo Angle Anomaly
}
\vspace{-5mm}
\author[a,b]{Andreas Crivellin,} 
\author[c]{Matthew Kirk,}
\author[d,e,f]{Teppei Kitahara,}
\author[c]{Federico Mescia}

\affiliation[a]{Physik-Institut, Universit\"at Z\"urich, Winterthurerstrasse 190, 8057 Z\"urich, Switzerland}
\affiliation[b]{Paul Scherrer Institut, 5232 Villigen PSI, Switzerland}
\affiliation[c]{Departament de F\'isica Quàntica i Astrof\'isica (FQA), Institut de Ciències del Cosmos (ICCUB), Universitat de Barcelona (UB), Spain} 
\affiliation[d]{IAR
\& 
KMI, Nagoya University, Nagoya 464–8602, Japan
}
\affiliation[e]{KEK Theory Center, IPNS, KEK, Tsukuba 305--0801, Japan}
\affiliation[f]{CAS Key Laboratory of Theoretical Physics, ITP,
CAS, Beijing 100190, China}

\emailAdd{andreas.crivellin@cern.ch}
\emailAdd{mjkirk@icc.ub.edu}
\emailAdd{teppeik@kmi.nagoya-u.ac.jp}
\emailAdd{mescia@ub.edu}
\medskip
\abstract{There are two tensions related to the Cabibbo angle of the CKM matrix. First, the determinations of $V_{us}$ from $K_{\mu 2}$, $K_{\ell3}$, and $\tau$ decays disagree at the $3\sigma$ level. Second, using the average of these results in combination with $\beta$ decays (including super-allowed $\beta$ decays and neutron decay), a deficit in first-row CKM unitarity with a significance of again about $3\sigma$ is found. These discrepancies, known as the Cabibbo Angle anomaly, can in principle be solved by modifications of $W$ boson couplings to quarks. However, due to $SU(2)_L$ invariance, $Z$ couplings to quarks are also modified and flavour changing neutral currents can occur. In order to consistently assess the agreement of a new physics hypothesis with data, we perform a combined analysis for all dimension-six Standard Model Effective Field Theory operators that generate modified $W$ couplings to first and second generation quarks.
We then study models with vector-like quarks, which are prime candidates for a corresponding UV completion as they can affect $W$-quark couplings at tree level, and we perform a global fit including flavour observables (in particular loop effects in $\Delta F=2$ processes). We find that the best fit can be obtained for the $SU(2)_L$ doublet vector-like quark $Q$ as it can generate right-handed $W$-$u$-$d$ and $W$-$u$-$s$ couplings as preferred by data.
} 

\begin{document} 
\maketitle
\allowdisplaybreaks
\renewcommand{\thefootnote}{\#\arabic{footnote}}
\setcounter{footnote}{0}

\section{Introduction}\label{sec:intro}

The Standard Model (SM) of particle physics has been very successfully tested and confirmed in the last decades with the Higgs discovery in 2012 providing the last missing constituent~\cite{ATLAS:2012yve,CMS:2012qbp}. As the Large Hadron Collider (LHC) at CERN has not (yet) found any new particles directly, precision experiments are becoming increasingly important to discover physics beyond the SM. In particular, an intriguing set of anomalies related to the violation of lepton flavour universality (see, e.g., Refs.~\cite{Crivellin:2021sff,Crivellin:2022qcj,Fischer:2021sqw, Artuso:2022ijh} for recent reviews) exist. 
\smallskip

Among them, there is the so-called Cabibbo Angle Anomaly (CAA)~\cite{Belfatto:2019swo,Grossman:2019bzp,Seng:2020wjq,Coutinho:2019aiy,Manzari:2020eum,Crivellin:2020ebi,Kirk:2020wdk,Crivellin:2020lzu,Capdevila:2020rrl,Belfatto:2021jhf,Crivellin:2021njn} with a significance of currently around the $3\sigma$ level~\cite{Bryman:2021teu,Cirigliano:2022yyo,Seng:2022ufm}.
The CAA consists of two tensions related to the determination of the Cabibbo angle: First, the different determinations of $|V_{us}|$ from $K_{\mu 2}$, $K_{\ell3}$, and $\tau$ decays disagree at the $3\sigma$ level. Second, using the average of these results in combination with $\beta$ decays, a deficit in first-row Cabibbo-Kobayashi-Maskawa (CKM) unitarity appears with a significance at the $3\sigma$ level. While the deficit in the first-row unitarity could be related to lepton-flavour-universality violating new physics (NP), see Refs.~\cite{Manzari:2021kma,Crivellin:2022ctt} for reviews, such a setup cannot solve the tensions between the different determinations of $|V_{us}|$. Intriguingly, however, both discrepancies (the CKM unitarity deficit and the tensions within $|V_{us}|$) could be explained via a modified $W$ couplings to quarks. 
\smallskip

Importantly, due to SU(2)$_L$ invariance, such a modified $W$ coupling to quarks in general leads to modified $Z$-quark-quark couplings as well, that enter electroweak precision observables, affect low-energy parity violation and can give contributions to the flavour changing neutral current (FCNC) processes. Therefore, a global fit is required to consistently assess the agreement of a specific NP scenario with data. The necessity of such a combined analysis becomes even more obvious when considering a UV complete model that can generate modified $W$ couplings to quarks. 
\smallskip

Here, we will study vector-like quarks (VLQs) as they give rise to such modifications already at tree level. While a new $4^{\rm th}$ generation of chiral fermions has been ruled out due to the combined constraints from LHC searches and flavour observables~\cite{Eberhardt:2012ck,Eberhardt:2012gv}, vector-like fermions can be added consistently without generating gauge anomalies. In fact, VLQs appear in many extensions of the SM such as grand unified theories~\cite{Hewett:1988xc,Langacker:1980js,delAguila:1982fs}, composite models or models with extra dimensions~\cite{Antoniadis:1990ew,Arkani-Hamed:1998cxo} and little Higgs models~\cite{Arkani-Hamed:2002ikv,Han:2003wu}. Furthermore, they have recently been studied intensively for phenomenological reasons since they can be considered part of the solution to $b\to s\ell^+\ell^-$ data~\cite{Altmannshofer:2014cfa,Gripaios:2015gra,Arnan:2016cpy,Arnan:2019uhr,Crivellin:2020oup,Bobeth:2016llm}, the tension in $(g-2)_\mu$~\cite{Czarnecki:2001pv,Kannike:2011ng,Dermisek:2013gta,Freitas:2014pua,Belanger:2015nma,Aboubrahim:2016xuz,Kowalska:2017iqv,Raby:2017igl,Choudhury:2017fuu,Calibbi:2018rzv,Crivellin:2018qmi,Capdevilla:2020qel,Capdevilla:2021rwo,Crivellin:2021rbq,Calibbi:2021pyh,Arcadi:2021glq,Paradisi:2022vqp, Allwicher:2021jkr} and the $W$ mass~\cite{Crivellin:2022fdf,Balkin:2022glu,Chowdhury:2022dps,Branco:2022gja} and are prime candidates for explaining the CAA~\cite{Belfatto:2019swo,Branco:2021vhs,Belfatto:2021jhf,Balaji:2021lpr}. In this case, not only the effect of modified $Z$ couplings to quarks, like in the effective field theory (EFT) case, must be taken into account, but also loop effects in flavour observables have to be included in a global analysis.
\smallskip

In this paper we will perform such a global analysis, first for the Standard Model Effective Field Theory (SMEFT), and then for models with VLQs coupling to first and second generation quarks.
We start by summarising the current status of the anomalies related to the Cabibbo angle in the next section. In \cref{sec:generic} we will describe the set up of our global fit, the matching of VLQs to the SMEFT, and discuss the relevant observables. Then in \cref{sec:eft} we use the global fit to analyse various EFT scenarios that correspond to modified gauge boson couplings to quarks, and see which scenarios provide the best fit to the current data. We then consider the different VLQ representations and their couplings to quarks in \cref{sec:vlqs} and conclude in \cref{sec:conclu}. Various useful results and further details are given in \cref{app:smeft_matching,app:BSM_ckm_values,app:smelli_obs,app:4d_scenarios}.
\medskip

\section{Current Status of Cabibbo Angle Anomaly}

\begin{figure}[t]
\begin{center}
\includegraphics[width=0.70\textwidth]{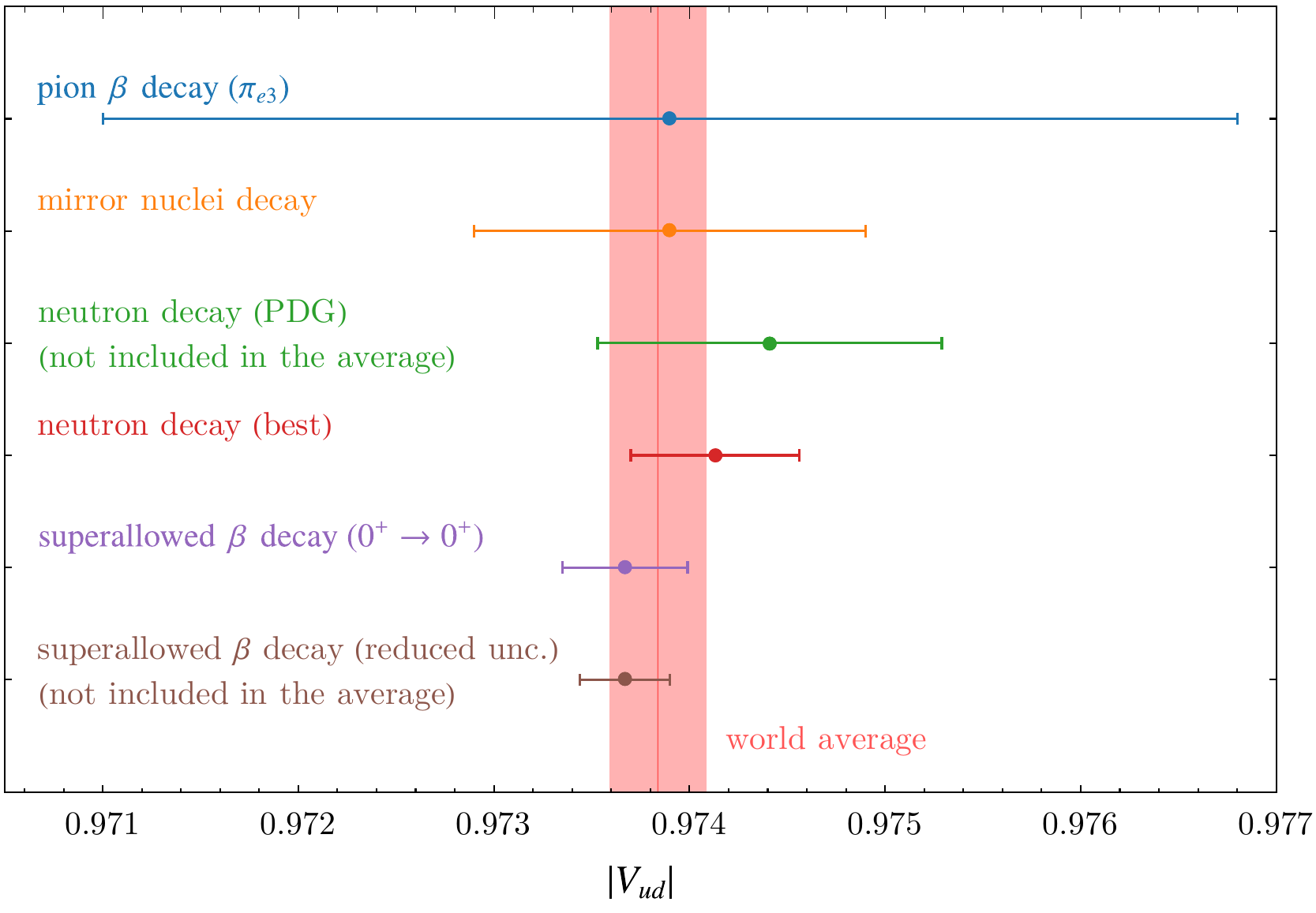}
\caption{Summary of the determinations of $|V_{ud}|$ from the various types of $\beta$ decays. The red band represents our world average~\eqref{eq:betaWA}.
The details of the extractions from
 neutron decay (best) and the super-allowed $\beta$ decays (reduced uncertainty) are given in the main text. Note that the pion $\beta$ decay, even though it is currently not competitive, is theoretically clean and will be strikingly improved by the PIONEER experiment~\cite{PIONEER:2022alm}.
}
\label{fig:beta_status}
\vspace{0.3cm}
\includegraphics[width=0.70\textwidth]{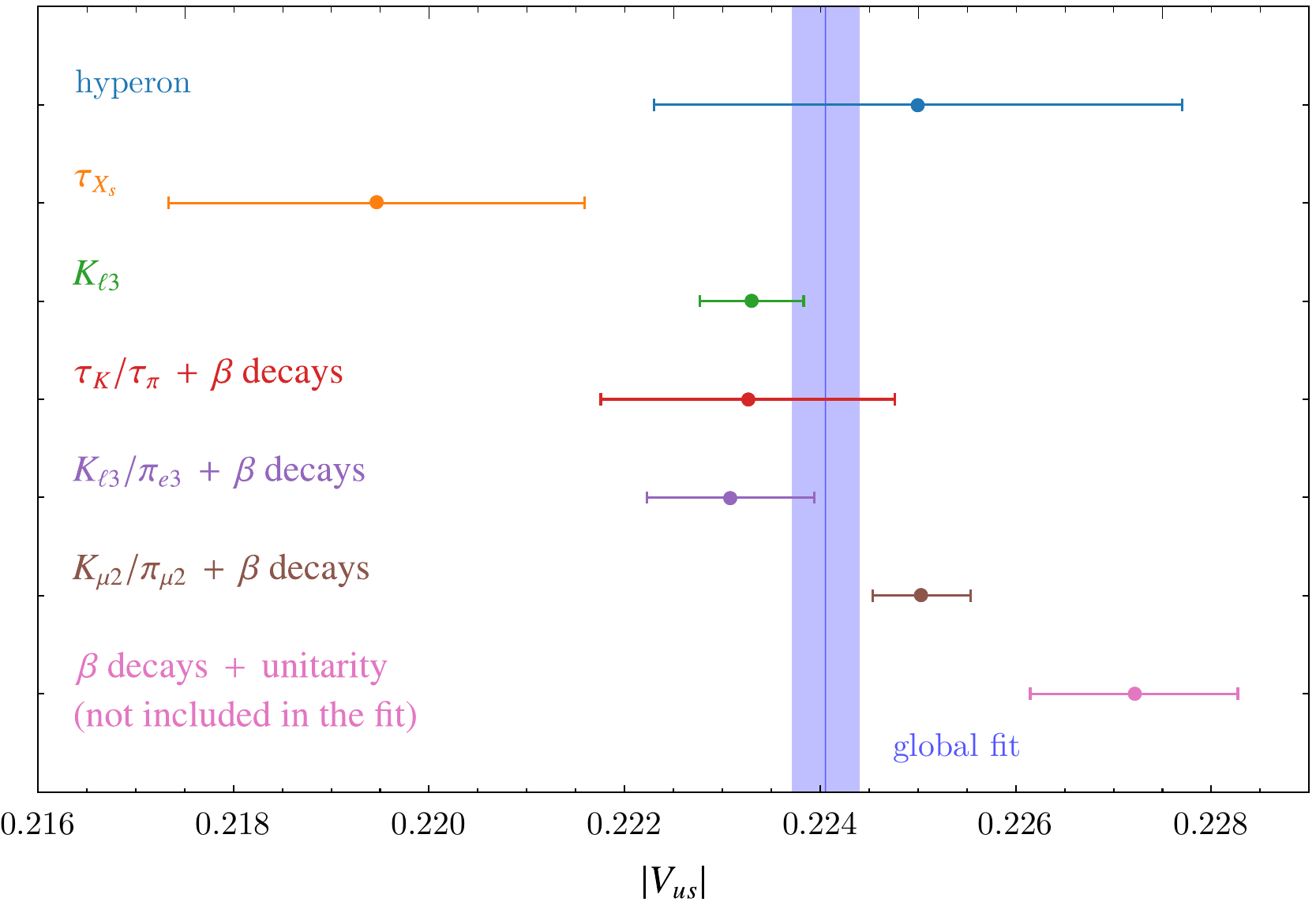}
\hspace{0.4cm}
\caption{Summary of the determinations of $|V_{us}|$ from various processes. The global fit value of $|V_{us}|$ is obtained in Eq.~\eqref{eq:Vusglobal}.
Note that the global fit does not include the CKM unitarity constraint.}
\label{fig:Vus_status}
\end{center}
\end{figure}

\begin{figure}[t]
\begin{center}
\includegraphics[width=0.8 \textwidth]{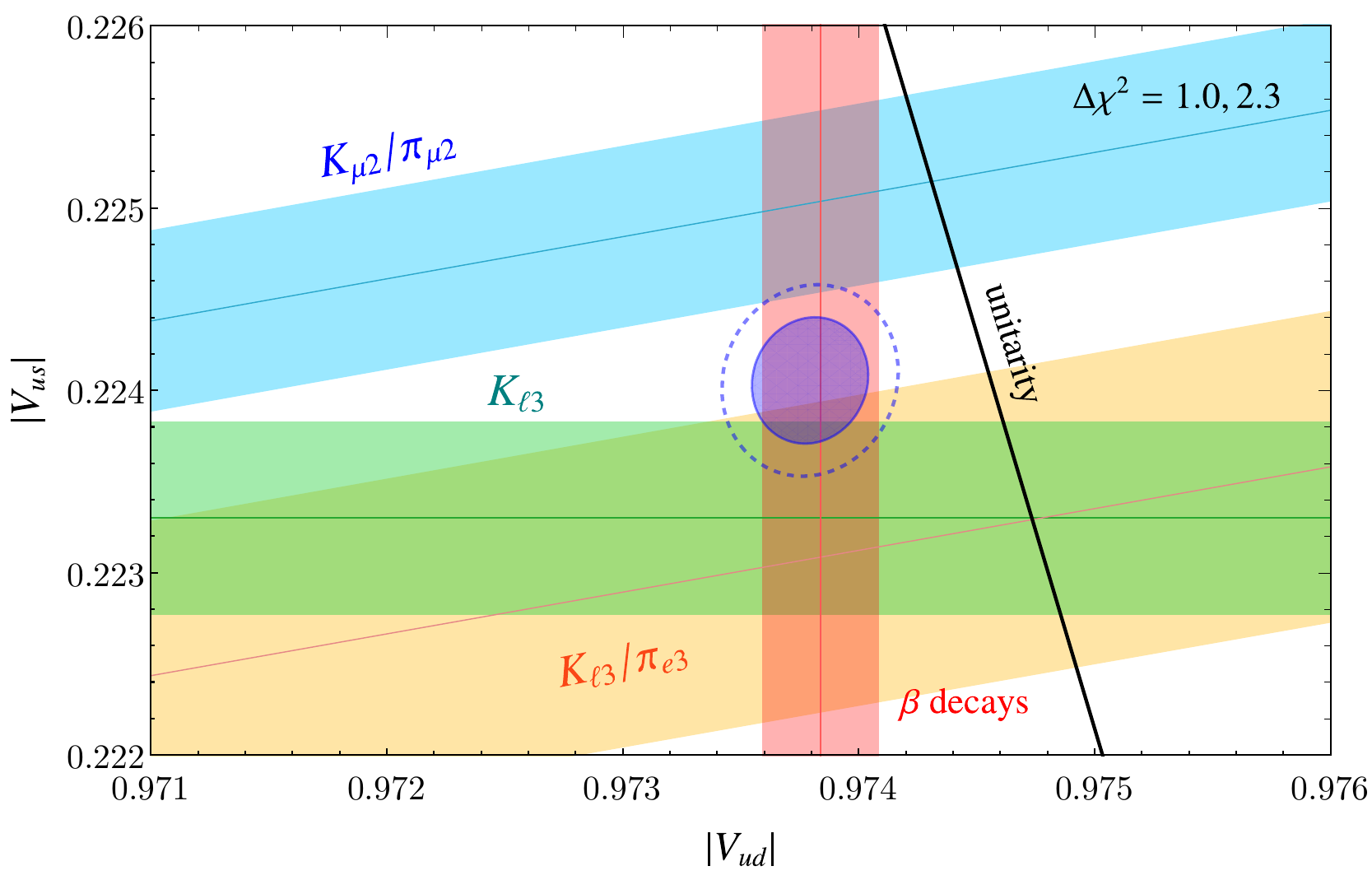}
\caption{
Global fit of all the available CKM determinations   
with $\Delta \chi^2 = 1$ (blue shaded) and $\Delta \chi^2 = 2.3$ (dashed circle). Only the $1\sigma$ regions from $\beta$ decays, $K_{\ell 3}$, $K_{\mu 2}/\pi_{\mu 2}$ and $K_{\ell 3}/\pi_{e 3}$ observables are shown. The black line represents the unitarity condition.
}
\label{fig:CAA_status_several}
\end{center}
\vspace{0.1cm}
\begin{center}
\includegraphics[width=0.8 \textwidth]{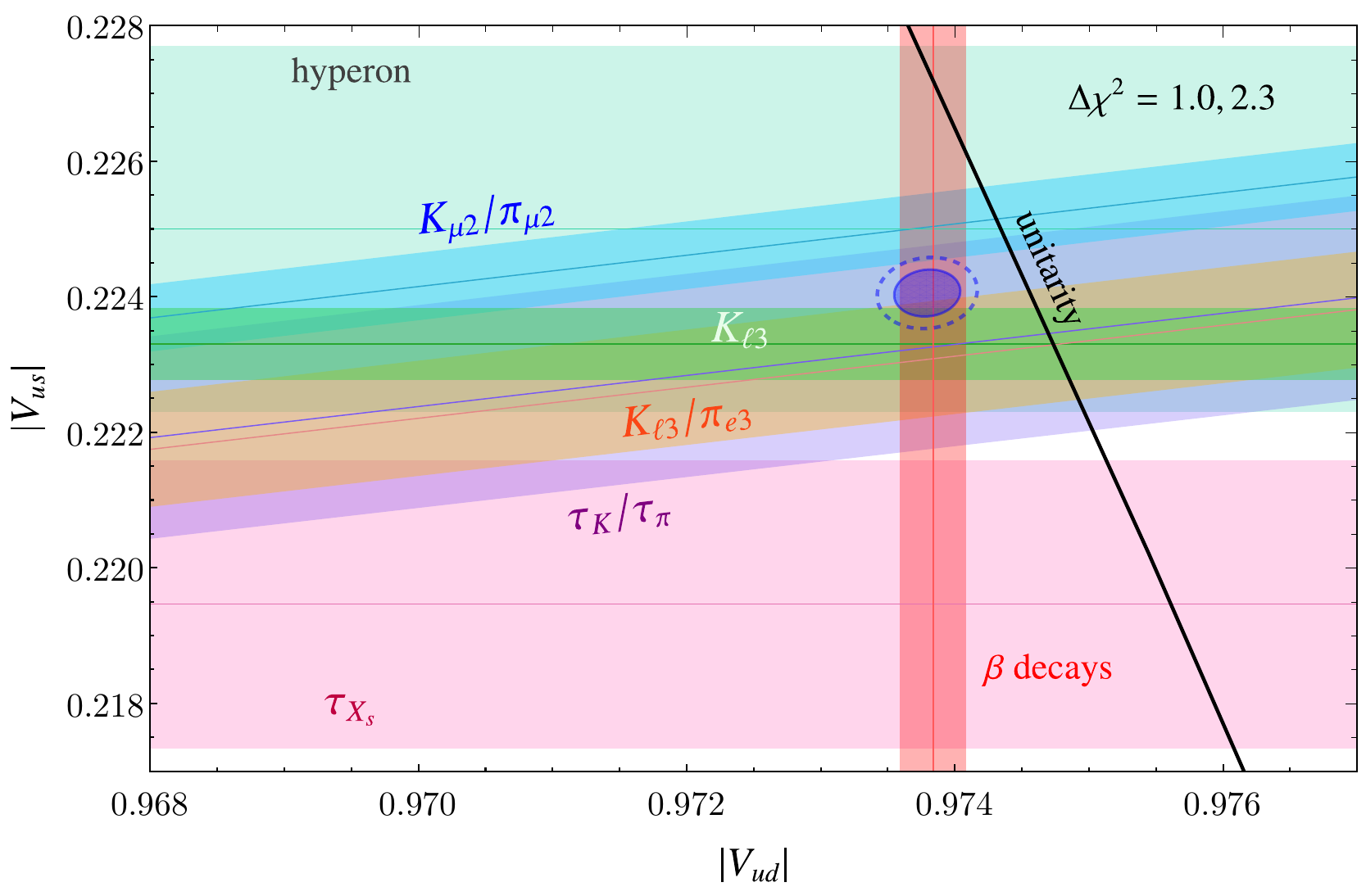}
\caption{Same as Fig.~\ref{fig:CAA_status_several},
but the $1\sigma$ regions of $|V_{us}|_{\rm hyperon} $, $ |V_{us}|_{\tau_{X_s}} $ and 
$|V_{us}/V_{ud}|_{\tau_K/\tau_\pi}$ observables are also shown. }
\label{fig:CAA_status}
\end{center}
\end{figure}

In this section, we review  the current situations of the $|V_{ud}|$ and $|V_{us}|$ determinations (which give rise to the CAA) summarized in Figs.~\ref{fig:beta_status}--\ref{fig:CAA_status} and Table~\ref{tab:CKM_determination}.
\smallskip

First, the CKM element $|V_{ud}|$ 
can be determined from various types of $\beta$ decays.
The latest determinations are $|\V{ud}|_{0^+\to 0^+}=0.97367(32)$ from the super-allowed $0^+ \to 0^+$ nuclear $\beta$ decay~\cite{Hardy:2020qwl,Cirigliano:2022yyo}, 
$|\V{ud}|_{n{\rm (PDG)}} = 0.97441(88)$ 
from the neutron decay~\cite{ParticleDataGroup:2022pth},
$|\V{ud}|_{\rm mirror} = 0.9739(10)$ 
from $\beta$ transitions of the mirror nuclei \cite{Hayen:2020cxh}, and 
$|\V{ud}|_{\pi_{e 3}} = 0.9739(29)$
from the pion $\beta$  decay ($\pi^+ \to \pi^0 e^+ \nu;~\pi_{e3}$)~\cite{Feng:2020zdc,Cirigliano:2022yyo}.
In these determinations, we use an estimation of Ref.~\cite{Cirigliano:2022yyo} for universal nuclear-independent radiative corrections from $\gamma W$-box diagrams $\Delta^V_R$ \cite{Marciano:2005ec} (see \cref{tab:theory_parameter_updates}). For the neutron decay,
it is known that the uncertainty of $|\V{ud}|_{n{\rm (PDG)}}$ is inflated by scale factors which come from inconsistencies in the data. By using the single most precise result for the neutron lifetime $\tau_n$ \cite{UCNt:2021pcg} and the nucleon isovector axial charge $g_A/g_V$ \cite{Markisch:2018ndu}, a better determination of  $|V_{ud}|_{n{\rm (best)}} = 0.97413(43)$ is possible~\cite{Cirigliano:2022yyo}. 
Combining $|V_{ud}|_{0^+\to0^+}$, $|V_{ud}|_{n{\rm (best)}} $, $|\V{ud}|_{\rm mirror}$, and $|\V{ud}|_{\pi_{e 3}}$, we obtain a weighted average of
\begin{align}
|V_{ud}|_{\beta}= 0.973\,84(25)\,.
\label{eq:betaWA}
\end{align}
Here, any correlation among systematic uncertainties of the radiative corrections is discarded, which should be a good approximation because uncertainties of $|V_{ud}|$ are dominated by the experimental one except for the super-allowed $\beta$ decays.\footnote{\label{footnote1}Note that the $|V_{ud}|$ determination is predominated by super-allowed $\beta$ decays where the largest uncertainty 
comes from nuclear-structure (NS) dependent radiative corrections  (corresponding to nuclear polarizability correction)~\cite{Gorchtein:2018fxl}, encoded in $\delta_{{\rm NS},E}$ in Ref.~\cite{Hardy:2020qwl}. Unfortunately, precise estimations of $\delta_{{\rm NS},E}$ are difficult~\cite{Seng:2022cnq} but the current value is considered to be very conservative~\cite{Gorchtein:2018fxl}.
Omitting the uncertainty from the $\delta_{{\rm NS},E}$ corrections (see \cref{tab:exp_parameter_updates}), 
$|\V{ud}|_{0^+\to 0^+{\rm (reduced~unc.)}}=0.97367(23)$ is obtained and 
the weighted average of the $\beta$ decays 
becomes 
$|\V{ud}|_{\beta\,{\rm (reduced~unc.)}}=
0.97378(20)$.}
\smallskip

\begin{table}[t]
\begin{tabularx}{\textwidth}{@{}LLXL@{}}
\toprule
\midrule
 & \text{Value} & \text{Observables} & 
\text{Label} \\
\midrule
\multirow{2}{*}{$V_{ud}$} & \multirow{2}{*}{\num{0.97367(32)}} & $Q$ value and lifetime $ft$ of super-allowed $0^+ \to 0^+$ nuclear
$\beta$ decays  & \multirow{2}{*}{$0^+ \!\to 0^+$}  \\
V_{ud} & \num{0.97413(43)} & $\tau_n~ \text{\cite{UCNt:2021pcg} ~and~} g_A/g_V~\text{\cite{Markisch:2018ndu}}$ & n\text{\,(best)} \\
V_{ud} & \num{0.97441(88)} & $\tau_n \text{~and~} g_A/g_V$ in PDG fit \cite{ParticleDataGroup:2022pth}& n\text{\,(PDG)}\\
V_{ud} & \num{0.9739(10)} & $ft$ of mirror nuclei decay 
& \text{mirror} \\
V_{ud} & \num{0.9739(29)} & 
$\Gamma(\pi^+ \to \pi^0 e^+ \nu)$ & 
\pi_{e3} \\
\midrule
V_{ud} & \num{0.97384(25)} & world average w/o $\pi_{e3}$ input & \beta\\
V_{ud} & \num{0.97384(25)} & world average  & \beta\\
\midrule
\midrule
V_{us} & \num{0.22330(53)} & $\Gamma(K_{S,L}\to\pi^- \ell^+ \nu), \,\Gamma(K^+\to\pi^0 \ell^+ \nu)$ & K_{\ell 3}  \\
V_{us} & \num{0.2195(21)} 
& $\Gamma(\tau \to X_s \nu)/\Gamma(\tau \to e \nu \bar\nu)$ & 
\tau_{X_s}  \\
\multirow{2}{*}{$V_{us}$} & \multirow{2}{*}{\num{0.2250(27)}} 
& $\Lambda^0 \to   p, \,\Sigma^- \to n,\, 
\Xi^- \to \Lambda^0,\, \Xi^0 \to \Sigma^+$  semi-leptonic decays & \multirow{2}{*}{\text{hyperon}}  \\
\midrule
V_{us} & \num{0.22314(51)} 
& world average&  K,\tau,\Lambda\\
\midrule
\midrule
V_{us}/V_{ud} & \num{0.23108(51)} & $\Gamma(K^+ \to \mu^+ \nu)/\Gamma(\pi^+ \to \mu^+ \nu)$& K_{\mu 2}/\pi_{\mu 2}  \\
V_{us}/V_{ud} & \num{0.22908(88)} & 
$\Gamma(K \to \pi \ell \nu)/\Gamma (\pi^+ \to \pi^0 e^+ \nu)$ & K_{\ell 3}/\pi_{e 3}  \\
V_{us}/V_{ud} & \num{0.2293(15)} & 
$\Gamma(\tau \to K^-  \nu)/\Gamma (\tau \to \pi^-  \nu)$ & 
\tau_K/\tau_\pi  \\
\midrule
V_{us}/V_{ud} & \num{0.23047(43)} &  world average  & {\rm ratios}  \\
\midrule
\midrule
V_{ud} & \num{0.97379(25)} &  global fit  & {\rm global}  \\
V_{us} & \num{0.22405(35)} &  global fit  & {\rm global}  \\
\midrule
\bottomrule
\end{tabularx}
\caption{Up-to-date extractions of the CKM elements needed to test the first-row unitarity.}
\label{tab:CKM_determination}
\end{table}

\begin{table}[t]
\begin{tabularx}{\textwidth}{@{}LLX@{}}
\toprule
\text{Parameter} & \text{Value} & Source \\
\midrule
f_+^{K^0 \to \pi^-}(0) & \num{0.9698(17)} & FLAG 2021 $N_f=2+1+1$ average, Eq.\,(76) in \cite{Aoki:2021kgd} \\
f_{K^\pm} / f_{\pi^\pm} & \num{1.1932(21)} & FLAG 2021 $N_f=2+1+1$ average, Eq.\,(81) in \cite{Aoki:2021kgd} \\
f_{K} / f_{\pi} & \num{1.1978(22)} & Isospin-limit $N_f=2+1+1$  average, Slide 34 of \cite{Moulson:CKM21}\\
\Delta_R^V \,(0^+\! \to0^+)& \num{0.02467(27)} & Average of \cite{Seng:2018yzq,Seng:2018qru,Czarnecki:2019mwq,Seng:2020wjq,Hayen:2020cxh,Shiells:2020fqp} from \cite{Cirigliano:2022yyo} \\
\delta_\text{EM}^{K^0 e}  \,(K_{\ell 3})& \num{0.0116(3)} & Table VI in \cite{Seng:2021wcf} \\
\delta_\text{EM}^{K^+ e}  \,(K_{\ell 3})& \num{0.0021(5)} & Table VI in  \cite{Seng:2021wcf} \\
\delta_\text{EM}^{K^0\mu} \,(K_{\ell 3})& \num{0.0154(4)} & Table IV in  \cite{Seng:2022wcw}\\
\delta_\text{EM}^{K^+\mu} \,(K_{\ell 3})& \num{0.0005(5)} & Table IV in  \cite{Seng:2022wcw}\\
\delta_\text{SU(2)}^{K^+ \pi^0} \,(K_{\ell 3})& \num{0.0252(11)} & Given in \cite{Cirigliano:2022yyo} as $\Delta_{\text{SU(2)}}$\\
\delta_\text{EM+SU(2)} (K_{\mu 2}/\pi_{\mu2}) & \num{-0.0126(14)} & Eq.\,(106) in \cite{DiCarlo:2019thl} \\
\bottomrule
\end{tabularx}
\caption{Updated values of the theoretical parameters which are used in this work. $\Delta^V_R$ is the universal nuclear-independent radiative correction to $\beta$ decays (see main text for more details), $\delta_\text{EM}$ are the electromagnetic corrections to $K_{\ell 3}$ decays (see references for details), $\delta_\text{SU(2)}$ is the isospin-breaking corrections to $K_{\ell 3}$ decays, and $\delta_\text{EM+SU(2)}$ is the difference in combined lattice calculations for electromagnetic and strong isospin-breaking corrections to $K_{\mu 2}$ and $\pi_{\mu 2}$.}
\label{tab:theory_parameter_updates}
\end{table}

Next, the matrix element $|V_{us}|$ can be determined from semi-leptonic decays of kaons and hyperons and from inclusive hadronic $\tau$ decays. By comparing theoretical predictions with data of the semi-leptonic kaon decays $K_{S,L} \to \pi^- \ell^+ \,\nu$ and $K^+ \to \pi^0 \ell^+ \,\nu$ with $\ell = e, \mu$ (labelled $K_{\ell 3}$), one can obtain~\cite{Cirigliano:2022yyo}, 
\begin{align}
    |V_{us}|_{K_{\ell 3}} = 0.223\, 30(53)\,,
\end{align}
where the latest evaluations of the long-distance electromagnetic (EM)  correction \cite{Seng:2021boy,Seng:2021wcf,Seng:2021nar,Seng:2022wcw}, the strong isospin-breaking
correction \cite{Cirigliano:2022yyo} (see \cref{tab:theory_parameter_updates}),
and the recent $K_S$ data from the KLOE-2 collaboration 
\cite{KLOE-2:2019rev,KLOE-2:2022dot} are used.
Here, we also used the FLAG 2021 $N_f=2+1+1$ value for $f_{+}^{K^0 \to \pi^-}(0)$,\footnote{
Lattice works contributing to the $f_{+}^{K^0 \to \pi^-}(0)$ FLAG average are in Refs.~\cite{Carrasco:2016kpy, FermilabLattice:2018zqv}.} and the form-factor parameters from Ref.~\cite{Moulson:CKM21}, see \cref{tab:theory_parameter_updates,tab:exp_parameter_updates}.
Beyond kaons, one can also use the hyperon semi-leptonic decays, $(\Lambda \to p, \Sigma \to n, \Xi \to \Lambda, \Xi \to \Sigma ) \,\ell\, \overline\nu$, which however lead to a slightly different yet less precise value~\cite{Cabibbo:2003ea,Geng:2009ik,ParticleDataGroup:2022pth}
\begin{align}
|V_{us}|_{\rm hyperon} = 0.2250(27)\,.
\end{align}

\begin{table}[t]
\begin{tabularx}{\textwidth}{@{}LLX@{}}
\toprule
\text{Parameter} & \text{Value} & Source \\
\midrule
\multirow{2}{*}{$\overline{\mathcal{F}t}\,(0^+ \to 0^+)$} & \SI{3072.24 \pm 1.85}{\s} & Eq.\,(22) in Hardy and Towner \cite{Hardy:2020qwl} \\
 & \SI{3072.24 \pm 1.21}{\s} & Hardy and Towner \cite{Hardy:2020qwl} without uncertainty of $\delta_{{\rm NS},E}$, Eq.\,(21) in \cite{Gorchtein:2018fxl} \\
\Lambda_+\,(K_{\ell 3}) & \num{25.55(38)e-3} & Slide 21 of \cite{Moulson:CKM21} \\
\ln C \,(K_{\ell 3})& \num{0.1992(78)} & Slide 21 of \cite{Moulson:CKM21} \\
\bottomrule
\end{tabularx}
\caption{Updated experimental inputs for the CKM determinations used in this work.}
\label{tab:exp_parameter_updates}
\end{table}

Inclusive hadronic $\tau$ decays also provide an opportunity to extract the matrix element $|V_{us}|$  by separating the strange and non-strange hadronic states.
Two representative determinations are reported: $ |V_{us}|_{\rm HFLAV}  = 0.2184 (21)$~\cite{Gamiz:2006xx,HFLAV:2022pwe,Lusiani:2022} and $ |V_{us}|_{\rm OPE+lattice}  = 0.2212(23)$~\cite{Hudspith:2017vew,Maltman:2019xeh}.
The former is based on the conventional operator product expansion (OPE) with using the vacuum saturation approximation~\cite{Pich:1999hc}, while the later is based on improved OPE series by fitting the lattice result~\cite{RBC:2010qam}.\footnote{%
Instead of the OPE approach, $|V_{us}|$ from the inclusive hadronic $\tau$ decays can be obtained based on the lattice-QCD simulation, where the spectral functions are evaluated by the lattice data of the hadronic vacuum polarization functions~\cite{RBC:2018uyk,Maltman:2019xeh,Aoki:2021kgd}.
This lattice-based determination provides 
$|V_{us}|_{\rm lattice}  = 0.2240 (18)$. Although this $|V_{us}|_{\rm lattice}$  is a little more accurate compared to the others, it does mostly rely on the $\tau \to K^- \nu $ data~\cite{RBC:2018uyk}, which is only $\sim 20\%$ of the inclusive strange-hadronic decays~\cite{HFLAV:2022pwe}.
These facts imply that this determination does not well represent the sum of the exclusive $\tau$ decays, as well as an unknown correlation with |\V{us}/\V{ud}| from exclusive $\tau$ decay ($\tau_K/\tau_\pi$ in Table~\ref{tab:CKM_determination}).
Therefore, we do not include this value in our analysis. }
Although they almost agree, 
there is no common consensus on which value, $|\V{us}|_{\rm HFLAV}$ or $|V_{us}|_{\rm lattice}$, to use~\cite{Lusiani} . Accordingly, we perform a weighted average of the two values
\begin{align}
\label{eq:inclusive_tau_combination}
 |V_{us}|_{\tau_{X_s}} =  0.2195(21) \,.
\end{align}
Here, a $100\%$ correlation of the statistical uncertainty and a naive average of systematics uncertainty are taken into account for simplicity because they are based on the same data. By using these $|\V{us}|$ determinations, we obtain  
a weighted average of $|V_{us}|_{K_{\ell 3}}$, $|V_{us}|_{\rm hyperon}$, and $|V_{us}|_{\tau_{X_s}}$,
\begin{align}
 |V_{us}|_{K,\tau,\Lambda} = 0.223\,14(51)\,.
\end{align}
These $|\V{us}|$ values are summarized in Fig.~\ref{fig:Vus_status} and Table~\ref{tab:CKM_determination}.
\smallskip

Third, the ratio $|V_{us}/V_{ud}|$ can be extracted from the several ratios of leptonic decay rates of kaon, pion  and $\tau$ leptons. The leptonic kaon-decay rate over the pion one, $K_{\mu2}/\pi_{\mu2}=\Gamma(K^+ \to \mu^+\nu)/\Gamma(\pi^+ \to \mu^+  \nu)$  provides~\cite{Moulson:CKM21}
\begin{align}
    \left| \frac{V_{us}}{V_{ud}}\right|_{K_{\mu2}/\pi_{\mu2}} = 0.231\,08(51)\,,
\end{align}
where the latest evaluation of the long-distance EM and strong isospin-breaking corrections \cite{Giusti:2017dwk,DiCarlo:2019thl} is used, see \cref{tab:theory_parameter_updates}.
Furthermore, the exclusive $\tau$-decay ratio $\Gamma(\tau \to K^- \nu)/\Gamma(\tau \to \pi^- \nu)$ (labelled by $\tau_K/\tau_\pi$) provides \cite{Lusiani:2022wcp}
 (see also Refs.~\cite{Arroyo-Urena:2021nil,Arroyo-Urena:2021dfe})
\begin{align}
    \left|\frac{V_{us}}{V_{ud}}\right|_{\tau_K/\tau_\pi} = 0.2293(15)\,.
\end{align}
In both cases, to avoid the double counting of strong isospin-breaking contribution, we have made use of the isospin-limit $N_f=2+1+1$ average for $f_{K} / f_{\pi}$, taken from Ref.~\cite{Moulson:CKM21}, see~\cref{tab:theory_parameter_updates}.\footnote{
The decay constant from the FLAG 2021~\cite{Aoki:2021kgd}, 
$f_{K^\pm} / f_{\pi^\pm}=1.1932(21)$, contains the strong isospin-breaking contribution in the average.}
\smallskip

In addition, it is recently pointed out in Ref.~\cite{Czarnecki:2019iwz} that the semi-leptonic kaon-decay rate over the pion $\beta$ decay, 
$\Gamma(K \to \pi \ell \nu)/\Gamma (\pi^+ \to \pi^0 e^+ \nu)$ (labelled by $K_{\ell 3}/\pi_{e3}$), provides~\cite{Seng:2021nar}
\begin{align}
  \left|\frac{V_{us}}{V_{ud}}\right|_{K_{\ell 3}/\pi_{e3}} = 0.229\,08(88) \,,
\end{align}
where the FLAG 2021 $N_f=2+1+1$ average for $f_{+}^{K^0 \to \pi^-}(0)$ is used.
Again, we obtain a weighted average of $|\V{us}/\V{ud}|_{K_{\mu2}/\pi_{\mu2}}$, 
$|\V{us}/\V{ud}|_{\tau_K/\tau_\pi}$ and $|\V{us}/\V{ud}|_{K_{\ell 3}/\pi_{e3}}$,
\begin{align}
    \left|\frac{V_{us}}{V_{ud}}\right|_{\rm ratios} = 0.230\,47(43)\,.
\end{align}
Here, a correlation via the form factor $f_K/f_\pi$ should be negligible because the uncertainty of $|\V{us}/\V{ud}|_{\tau_K/\tau_\pi}$ is dominated by the experimental data.
\smallskip

Finally, we perform a global analysis within the SM.
In Figs~\ref{fig:CAA_status_several} and \ref{fig:CAA_status}, the global fit result including $|V_{ud}|_{\beta}$, 
$ |V_{us}|_{K,\tau,\Lambda} $ and 
$|V_{us}/V_{ud}|_{\rm ratios} $ is shown by the blue circles.
In Fig.~\ref{fig:CAA_status_several}, only $\beta$ decays, $K_{\ell 3}$, $K_{\mu 2}/\pi_{\mu 2}$ and $K_{\ell 3}/\pi_{e 3}$ are displayed (but all data are included in the global fit), while Fig.~\ref{fig:CAA_status} shows all data.
The black line stands for the unitarity condition: $|V_{ud}|^2+|V_{us}|^2 + |V_{ub}|^2=1$ with $|V_{ub}| \approx 0.00377$ (from~\cite{Charles:2004jd}, one could also use \cite{UTfit:2022hsi}, however the actual value is irrelevant due to its smallness).
The blue shaded circle corresponds to $\Delta \chi^2 \leq 1$, while the dashed circle is $\Delta \chi^2 = 2.3$.
In the $\chi^2$ analysis, we included a correlation between $K_{\ell 3}$ and  $K_{\ell 3}/\pi_{e 3}$ because they share the same kaon data and common form factor $f_{+}^{K^0 \to \pi^-}(0)$. We set $100\%$ correlation for these common uncertainties. 
\smallskip

Our global fit results are
\begin{align}
    |V_{ud}|_{\text{global}} & = 0.973\,79 (25)\,,
    \label{eq:Vudglobal}\\
    |V_{us}|_{\text{global}} & = 0.224\,05 (35)\,,
    \label{eq:Vusglobal}
\end{align}
and
\begin{align}
    \Delta_{\text{CKM}}^{\text{global}} & \equiv | V_{ud} |^2_{\text{global}} + | V_{us}|^2_{\text{global}} + |V_{ub}|^2 -1  = -0.001\,51(53)\,,
    \label{eq:CAAglobal}
\end{align}
with a $ |V_{ud}|_{\text{global}}$--$|V_{us}|_{\text{global}}$ correlation of $0.09$. This $\Delta_{\text{CKM}}^{\text{global}}$
 implies $- 2.8\sigma$ deviation from the unitarity condition of the CKM matrix.\footnote{If one ignores the large uncertainty from the nuclear-structure dependent corrections to the super-allowed $\beta$ decays and use 
$|\V{ud}|_{\beta\,{\rm (reduced~unc.)}}$ (see footnote\,\ref{footnote1}), 
a $- 3.7\sigma$ deviation from unitarity is observed in the global fit.}
\smallskip

 Also, one can define different CKM unitarity tests 
 by taking each pair of the best measurements ($\beta$ and the kaon decays) individually~\cite{Cirigliano:2022yyo}, which could distinguish each NP scenario, 
\begin{alignat}{2}
\begin{aligned}
\Delta_{\text{CKM}}^{(1)} &\equiv | V_{ud}|^2_{\beta}  + |V_{us}|^2_{K_{\ell 3}} + |V_{ub}|^2 -1 & =& - 0.001\,76 (54) \,, \\
\Delta_{\text{CKM}}^{(2)} & \equiv | V_{ud}|^2_{\beta}  + |V_{us}|^2_{K_{\mu 2}/\pi_{\mu 2},\,\beta} + |V_{ub}|^2 -1~ & =& - 0.000\,98 (56) \,, \\
\Delta_{\text{CKM}}^{(3)} &\equiv | V_{ud}|^2_{K_{\mu 2}/\pi_{\mu 2},\, K_{\ell 3}}  + |V_{us}|^2_{K_{\ell 3}} + |V_{ub}|^2 -1 & =& - 0.0163 (62)\,,
\label{eq:testunitarity}
\end{aligned}
\end{alignat}
corresponding to $-3.3\sigma,\, -1.8\sigma,\,-2.6\sigma$ discrepancies, respectively.

We summarize the determinations of $|V_{us}|$ from various observables in Fig.~\ref{fig:Vus_status}. There, the blue band represents the global fit of $|V_{us}|$ in Eq.~\eqref{eq:Vusglobal} in which the CKM unitarity condition is not included. It is shown that $|V_{us}|_{\tau_{X_s}}$ (orange bar) is a little bit smaller than the other determinations; $3.3\sigma,\,2.6\sigma,\,1.8\sigma$ discrepancies 
by comparing to $\beta$ decays with unitarity (magenta), 
$K_{\mu 2}/\pi_{\mu 2}$ with $\beta$ decays (brown), and $K_{\ell 3}$ (green), respectively. 
\smallskip

Before closing this section, we give a brief summary of the status of {\em first-column} CKM unitarity, i.e.,
$|\V{ud}|^2 + |\V{cd}|^2 + |\V{td}|^2 =1$.
The CKM element $|\V{cd}|$ can be determined from leptonic and semi-leptonic $D$-meson decays and by a charmed-hadron production via neutrino-nucleon scattering~\cite{CHORUS:2008vjb}.
The world average is
$|\V{cd}| = 0.221 (4)$~\cite{ParticleDataGroup:2022pth}, 
which is dominated by $D^+ \to \mu^+ \nu$~\cite{BESIII:2013iro}. The element $|\V{td}|$ can be determined indirectly by global CKM fit. 
The current world average is 
$|\V{td}| = 0.0086 (2)$~\cite{ParticleDataGroup:2022pth}, however, the actual value is irrelevant due to its smallness.
Combining this in a global fit with $|V_{ud}|$ of Eq.~\eqref{eq:Vudglobal}, we find the first-column CKM unitarity
\begin{align}
  \Delta_{\rm CKM}^{1^{\rm st}{\rm column}} \, \equiv  |V_{ud}|^2_{\rm global} + |V_{cd}|^2 + |V_{td}|^2 - 1 = -0.0028(18)\,,
\end{align}
implying deviation of $1.5\sigma$. 
Here, the statistical uncertainty of $D^+ \to \mu^+ \nu$~\cite{HFLAV:2022pwe} dominates, which will be reduced by the Belle~II~\cite{Belle-II:2018jsg} and BES III experiments~\cite{BESIII:2020nme}.
\medskip

\section{Setup}
\label{sec:generic}

In this section we first establish our conventions for the SMEFT and the extensions of the SM by VLQs. We then discuss our fit method and the most important constraints used in the global analysis.
\smallskip

\subsection{SMEFT}
We write the SMEFT Lagrangian as
\begin{equation}
\Lagrangian_\text{SMEFT} = \mathcal{L}_{\rm SM} + \sum_i C_i Q_i\,,
\end{equation}
such that the SMEFT coefficients have dimensions of inverse mass squared. We use the Warsaw basis~\cite{Grzadkowski:2010es}, as well as the corresponding conventions, in which the operators generating modified gauge-boson couplings to quarks (at tree-level) are given by
	\begin{align}
			Q_{H q}^{(1)ij} &= (H^{\dagger}i\overset{\leftrightarrow}{D_{\mu}}H)(\bar{q}_i\gamma^{\mu} P_L q_j)\,, &
			Q_{H q}^{(3)ij} &= (H^{\dagger}i\overset{\leftrightarrow}{D_{\mu}^I}H)(\bar{q}_i\tau^I\gamma^{\mu} P_L q_j)\,,\notag\\
			Q_{H u}^{ij} &= (H^{\dagger}i\overset{\leftrightarrow}{D_{\mu}}H)(\bar{u}_i\gamma^{\mu}P_R u_j)\,, &
			Q_{H d}^{ij} &= (H^{\dagger}i\overset{\leftrightarrow}{D_{\mu}}H)(\bar{d}_i\gamma^{\mu}P_R d_j)\,,\notag\\
			Q_{H ud}^{ij} &= i(\tilde{H}^{\dagger}D_{\mu}H)(\bar{u}_i\gamma^{\mu} P_R d_j)\,. & &
		\label{eq:HiggsqOperators}
	\end{align}
We work in the down-basis such that CKM elements appear in transitions involving left-handed up-type quarks after electroweak (EW) symmetry breaking. This means we write the left-handed quarks doublet as $q^T_i = \begin{pmatrix}(V^\dag u_{L})_{i} & d_{L, i} \end{pmatrix}$, where $V$ is the CKM matrix. With this conventions, the modified $W$ and $Z$ couplings are given by
\begin{equation}
\Lagrangian_{W,Z} = 
\begin{aligned}[t]
&- \frac{g_2}{\sqrt{2}} W^+_\mu \, \bar{u}_i \gamma^\mu \left( \left[ V \cdot \left( \mathbbm{1} + v^2  \CHqthree \right) \right]_{ij} P_L + \frac{v^2}{2} \CHudbracket_{ij} P_R \right) d_j + \hc \\
&- \frac{g_2}{6 c_W} Z_\mu \, \bar{u}_i \gamma^\mu 
	\begin{aligned}[t]
	\Bigg( &\left[ (3-4s_W^2) \mathbbm{1} + 3 v^2 \, V \cdot \left\{ \CHqthree - \CHqone \right\} \cdot V^\dagger \right]_{ij} P_L \\
	- &\left[4s_W^2 \mathbbm{1} + 3 v^2 C_{Hu}\right]_{ij} P_R \Bigg) u_j
	\end{aligned} \\
&- \frac{g_2}{6 c_W} Z_\mu \, \bar{d}_i \gamma^\mu 
	\begin{aligned}[t]
 	\Bigg( &\left[ (2s_W^2-3) \mathbbm{1} - 3 v^2 \left\{ \CHqthree + \CHqone \right\} \right]_{ij} P_L \\
	+ &\left[ 2s_W^2 \mathbbm{1} - 3 v^2 C_{Hd}\right]_{ij} P_R \Bigg) d_j \,,
	\end{aligned}
\end{aligned}
\label{eq:WZCH}
\end{equation}
where $v \approx \SI{246}{\GeV}$.

\subsection{Vector-Like Quarks}

There are seven possible VLQs that can mix with SM quarks after EW symmetry breaking:
\begin{equation}
\begin{aligned}
U&: \QuantumNumbers{3}{1}{\sfrac{2}{3}} \,, &D &: \QuantumNumbers{3}{1}{-\sfrac{1}{3}} \,, &Q &: \QuantumNumbers{3}{2}{\sfrac{1}{6}} \,, \\
Q_5 &: \QuantumNumbers{3}{2}{-\sfrac{5}{6}} \,, &Q_7 &: \QuantumNumbers{3}{2}{\sfrac{7}{6}} \,, \\
T_1 &: \QuantumNumbers{3}{3}{-\sfrac{1}{3}} \,, &T_2 &: \QuantumNumbers{3}{3}{\sfrac{2}{3}} \,.
\end{aligned}
\end{equation}
The numbers in the brackets denote the representation under the SM gauge group $SU(3)\times SU(2)_L\times U(1)_Y$. The Lagrangian describing their interactions with the Higgs and SM quarks is 
\begin{align}
-\mathcal{L}_{\rm VLQ} =
&\xi_{fi}^{U}\bar{U}_{f}\tilde{H}^{\dagger}q_{i} 
+ \xi_{fi}^{D}\bar{D}_{f}H^{\dagger}q_{i} 
+\xi_{fi}^{u}\bar{Q}_{f}\tilde{H}u_{i} +\xi_{fi}^{d}\bar{Q}_{f}Hd_{i}
\\
+ &\xi_{fi}^{Q_{5}}\bar{Q}_{5,f}\tilde{H}d_{i}+\xi_{fi}^{Q_{7}}\bar{Q}_{7,f}Hu_{i} +\frac{1}{2}\xi_{fi}^{T_{1}}H^{\dagger}\tau\cdot\bar{T}_{1,f}q_{i}+\frac{1}{2}\xi_{fi}^{T_{2}}\tilde{H}^{\dagger}\tau\cdot\bar{T}_{2,f}q_{i} + \text{h.c.}\,,\nonumber
\end{align}
where $q$ is the left-handed quark doublet, $u,d$ are the right handed quark singlets, and $i$ and $f$ are flavour indices for the SM quarks and new VLQs, respectively. Note that therefore $f$ does not necessarily need to run from 1 to 3 as the number of generations of VLQs is arbitrary (i.e.~unknown). We disregard possible couplings between two VLQs representations and the SM Higgs as they are not relevant (at the dimension-six level) for the modification of gauge boson couplings to quarks.
\smallskip

With these conventions, the matching obtained by integrating out the VLQs at tree level onto the SMEFT is 
\begin{equation}
\label{eq:VLQ_SMEFT_tree_matching}
\begin{aligned}
\left[C_{Hu}\right]_{ij} &= -\frac{\xi^{u}_{fj} \xi^{u *}_{fi}}{2M_{Q_{f}}^2} + \frac{\xi^{Q_7}_{fj} \xi^{Q_7 *}_{fi}}{2M_{Q_{7f}}^2} \,, \\
\left[C_{Hd}\right]_{ij} &= \frac{\xi^{d}_{fj} \xi^{d *}_{fi}}{2M_{Q_{f}}^2} - \frac{\xi^{Q_5}_{fj} \xi^{Q_5 *}_{fi}}{2M_{Q_{5f}}^2} \,, \\
\left[C_{Hud}\right]_{ij} &= \frac{\xi^{d}_{fj} \xi^{u *}_{fi}}{M_{Q_{f}}^2} \,, \\
\CHqonebracket_{ij} &= \frac{\xi^{U}_{fj} \xi^{U*}_{fi}}{4M_{U_f}^2} - \frac{\xi^{D}_{fj} \xi^{D*}_{fi}}{4M_{D_f}^2} - \frac{3\xi^{T_1}_{fj} \xi^{T_1*}_{fi}}{16M_{T_{1f}}^2} + \frac{3\xi^{T_2}_{fj} \xi^{T_2*}_{fi}}{16M_{T_{2f}}^2} \,, \\
\CHqthreebracket_{ij} &= -\frac{\xi^{U}_{fj} \xi^{U*}_{fi}}{4M_{U_f}^2} - \frac{\xi^{D}_{fj} \xi^{D*}_{fi}}{4M_{D_f}^2} + \frac{\xi^{T_1}_{fj} \xi^{T_1*}_{fi}}{16M_{T_{1f}}^2} + \frac{\xi^{T_2}_{fj} \xi^{T_2*}_{fi}}{16M_{T_{2f}}^2} \,.
\end{aligned}
\end{equation}
Note that $C_{uH}$ and $C_{dH}$ are also generated at tree-level, but as being proportional to the tiny masses of first and second generation quarks, they are not relevant for our study (if the couplings of the Higgs to two different VLQs are neglected). In Higgs decays, this suppression is removed when normalising to the SM rate, however these decays have not been measured (nor are they expected to be in the near future), and for flavour observables the quark mass suppression compared to the effects from the modified gauge boson couplings is restored. At one-loop, there are also contributions to the $\Delta F=2$ operators $Q_{qq}^{(1,3)}$, $Q_{qu}^{(1,8)}$, and $C_{uu}$ which affect \Dz--\Dzb  and kaon mixing and give rise to relevant bounds.%
The full expressions for the related Wilson coefficients can be found in \cref{app:smeft_matching}.
\footnote{At one-loop, the Wilson coefficient of SMEFT operator $Q_{HD}$, which modifies the $W$ mass, is generated. The latest results from CDF II~\cite{CDF:2022hxs}, which hint at a sizable deviation from the SM prediction, could be explained by VLQs with large couplings, i.e., bigger than one. However, we do not consider this possibility here and therefore do not include the measurement in the global fit. For the interested reader, a recent study of the $W$ mass in VLQ models has been performed in Ref.~\cite{Belfatto:2023tbv}.}

\medskip

\subsection{Fit Method and Observables}
\label{sef:fitmethod}

We use \smelli \texttt{v2.3.2}~\cite{Aebischer:2018iyb,smelli_2_3_2} (which is built on \flavio~\cite{Straub:2018kue} for the observable calculations, and \wilson~\cite{Aebischer:2018bkb} for the RG evolution in the SMEFT and the low-energy EFT (LEFT)) for our global fit. To efficiently sample the likelihood in our scenarios with more than two free parameters, we use an MCMC library \program{PyMultiNest}~\cite{Buchner:2014nha,Feroz:2008xx,Feroz:2013hea} with the software package \program{corner}~\cite{Foreman-Mackey2016} for visualisation.
\smallskip

Since our NP effects change the extraction of the CKM elements, the theory predictions of CKM dependent observables are non-trivial and a consistent treatment is necessary.
Following Ref.~\cite{Descotes-Genon:2018foz} we determine the Cabibbo angle, at each parameter point in parameters space using $K_{\mu2}/\pi_{\mu2}$ as input and take into account the NP effects, and then calculate \V{ud} and \V{us} using the unitarity of the CKM matrix of the SM Lagrangian.\footnote{Note that choosing $K_{\mu2}/\pi_{\mu2}$ to determine the Cabibbo angle is arbitrary in the sense that any other determination could be used and the final result of the global fit does not depend on this choice of the input scheme.} This then fixes the theory parameters necessary for the calculation of the other observables that depend on CKM elements which are then compared to their measured values when performing the fit.
\smallskip

In our fit we include all $\beta$ decays, along with $K_{\ell 3}$. The two exclusive $\tau$ decays $\tau \to \pi \nu$ and $\tau \to K \nu$ are included separately, rather than as a single ratio. We also include charged-current $D$ decays (since these are strongly sensitive to $\V{cd} \approx -\V{us}$), with both total branching ratios and individual $q^2$-binned data.\footnote{Note that we added these manually, since they are not included by default in \smelli \texttt{v2.3.2}, but they will be included in a future public release of \smelli.} Furthermore, in the later figures we refer to a single ``CKM'' region, this means the region in which all the different charged-current observables (listed explicitly in \cref{tab:smelli_Kl3_obs,tab:smelli_beta+n_obs,tab:smelli_Ddecay_obs,tab:smelli_tau_excl_obs} and Eq.~\eqref{eq:inclusive_tau_combination}) are in best agreement with data. Note also that this extraction of the CKM elements, is also used later in the calculation of the SM prediction for CP violation in kaon mixing ($\epsilon_K$).
\smallskip

The most relevant observables already contained within \smelli, which we updated with our input (see \cref{tab:theory_parameter_updates,tab:exp_parameter_updates}), are listed in \cref{app:smelli_obs}. Concerning kaon FCNC observables, both $\Delta S=2$ ($\epsilon_K$) and $\Delta S=1$ ($K \to \pi \nu \nu$  and $K \to \ell^+ \ell^-$) are included. Specifically concerning $\epsilon_K$, using our input parameters, \flavio gives a SM prediction for $\epsilon_K$ of \num{2.12 \pm 0.32 e-3}. Compared to the prediction in Ref.~\cite{Brod:2022har}, we have an \SI{80}{\percent} larger error, which can be mainly attributed to larger CKM uncertainties due to our BSM CKM treatment described above. It has previously been shown in Refs.~\cite{Endo:2016tnu,Bobeth:2017xry,Endo:2018gdn} that the dominant NP contribution to $\epsilon_K$ comes through diagrams with a $Z$-$s$-$d$ on one side, and a SM one-loop correction on the other, which leads to enhanced sensitivity to right-handed $Z$-$s$-$d$ couplings. In our fit, these effects are taken into account through the one-loop matching of the SMEFT onto the LEFT, as implemented in \wilson. Finally, for the effects in \Dz--\Dzb mixing, we include the one-loop induced $\Delta F=2$ coefficients, along with contributions from two insertions of the $\Delta F=1$ modified $Z$ couplings, which are formally of dimension-eight in the SMEFT power counting.
However, since a reliable SM prediction for $\Delta M_D$ is still unavailable, to be conservative (and also in light of our partial inclusion of dimension-eight SMEFT effects) we use a Gaussian likelihood for the NP contribution with mean 0 and standard deviation equal to the current experimental central value~\cite{HFLAV:2022pwe,hflav-charm-2021}.
\smallskip

In addition to these observables already present in \smelli, we implemented low-energy parity violation in \flavio, based on Ref.~\cite{Crivellin:2021bkd}, which can provide similarly strong bounds on VLQs as electroweak precision measurements. For this we added to the likelihood a contribution which comes from the $Q_\text{weak}$ experiment~\cite{Qweak:2018tjf} and the measurement of atomic parity violation in ${}^{133} \text{Cs}$~\cite{Cadeddu:2021dqx,Wood:1997zq,Guena:2004sq}. We also include a contribution from inclusive $\tau$ decays, based on our combination detailed above.
\medskip

\section{Analysis and Results}\label{sec:analysis}

We now perform our global analysis with the method and observables discussed in the last section. We start with the SMEFT where we use the Wilson coefficients as input at a scale of \SI{1}{\TeV} and evolve them to the scale of the observables, while for the VLQs we consider a matching scale of \SI{2}{\TeV}.
\smallskip

\subsection{SMEFT results for modified gauge boson couplings}
\label{sec:eft}

First of all, according to \eq{eq:WZCH}, while both $\CHqthreebracket_{11}$ (which generates a modification of the left-handed $W$-$u$-$d$ coupling) and $\CHudbracket_{11}$ (which generates a right-handed $W$-$u$-$d$ coupling) can in principle explain the deficit in first-row CKM unitarity, the disagreement between $V_{us}$ from $K_{\mu 2}$, $K_{\ell3}$ and $\tau$ decays can only be accounted for by $\CHudbracket_{12}$, i.e.~a right-handed $W$-$u$-$s$ coupling is necessary~\cite{Bernard:2007cf} (see appendix~\ref{app:BSM_ckm_values} for details). Therefore, we will focus on scenarios with these coefficients in the following.
\smallskip

\paragraph{1-D scenarios}

\begin{figure}
\includegraphics[width=\textwidth]{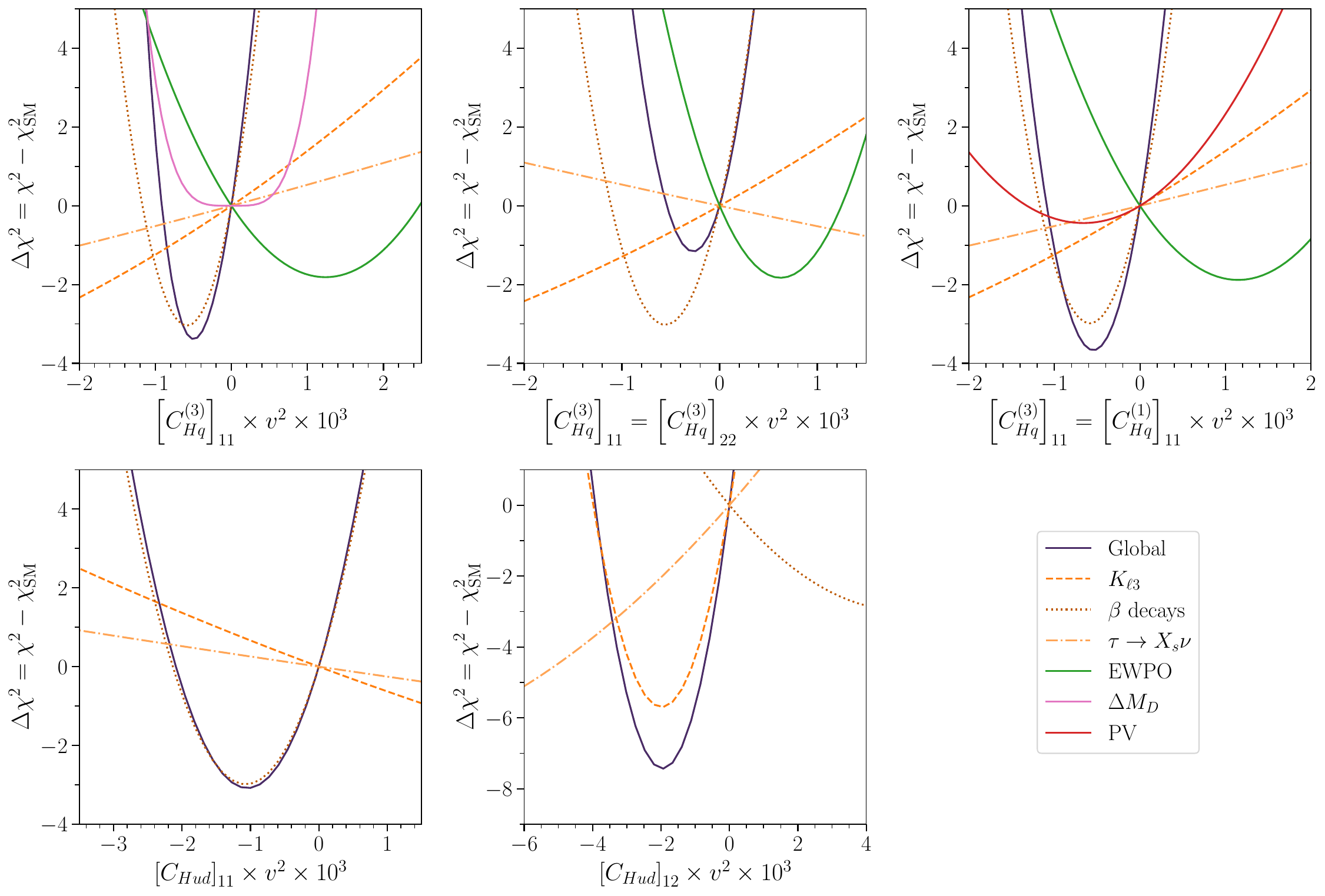}
\caption{$\Delta \chi^2$ w.r.t.~the SM as a function of the values of the Wilson coefficients for the 1D scenarios considered in the main text.}
\label{fig:eft_1d_scenarios}
\end{figure}

First, we consider a non-zero value of the Wilson coefficient $\CHqthreebracket_{11}$, where from our global fit we find
\begin{equation}
\CHqthreebracket_{11} \times v^2 = \num{-0.50 +- 0.25 e-3}\,.
\end{equation}
As we are working in the down-basis, no constraints from kaon physics arise, however, CKM rotations lead to effects in \Dz--\Dzb mixing, which are (despite our very conservative bound and the fact that it is a dimension-eight effect) stronger than the electroweak precision observables (EWPO) (see top-left panel in \cref{fig:eft_1d_scenarios}). However, the bounds from \Dz--\Dzb mixing can be weakened or avoided by using a flavour structure that respects $U(2)$ flavour ($\CHqthreebracket_{11} = \CHqthreebracket_{22}$) or by cancelling the effect in $Z$ couplings to up quarks via $\CHqthreebracket_{11} = \CHqonebracket_{11}$, respectively. For these two scenarios, shown in the top-middle and top-right panel of \cref{fig:eft_1d_scenarios}, we find
\begin{align}
\CHqthreebracket_{11} = \CHqthreebracket_{22}\times v^2 &= \num{-0.27 +- 0.25 e-3}\,,\\
\CHqthreebracket_{11} = \CHqonebracket_{11}  \times v^2 &= \num{-0.55 +- 0.28 e-3}\,.
\end{align}
Considering instead modifications of the right-handed $W$-$u$-$d$ or $W$-$u$-$s$ vertex, no effects in \Dz--\Dzb mixing and $Z$-pole observables arise (see bottom panels in \cref{fig:eft_1d_scenarios})
and we find
\begin{align}
\CHudbracket_{11}  \times v^2 &= \num{-1.0 \pm 0.6 e-3}\,,\\
\CHudbracket_{12}  \times v^2 &= \num{-2.0 \pm 0.7 e-3}\,,
\end{align}
\smallskip
The corresponding pulls for all scenarios are given in Table~\ref{tab:eft_best_fits}.
\medskip

\begin{figure}
\includegraphics[width=\textwidth]{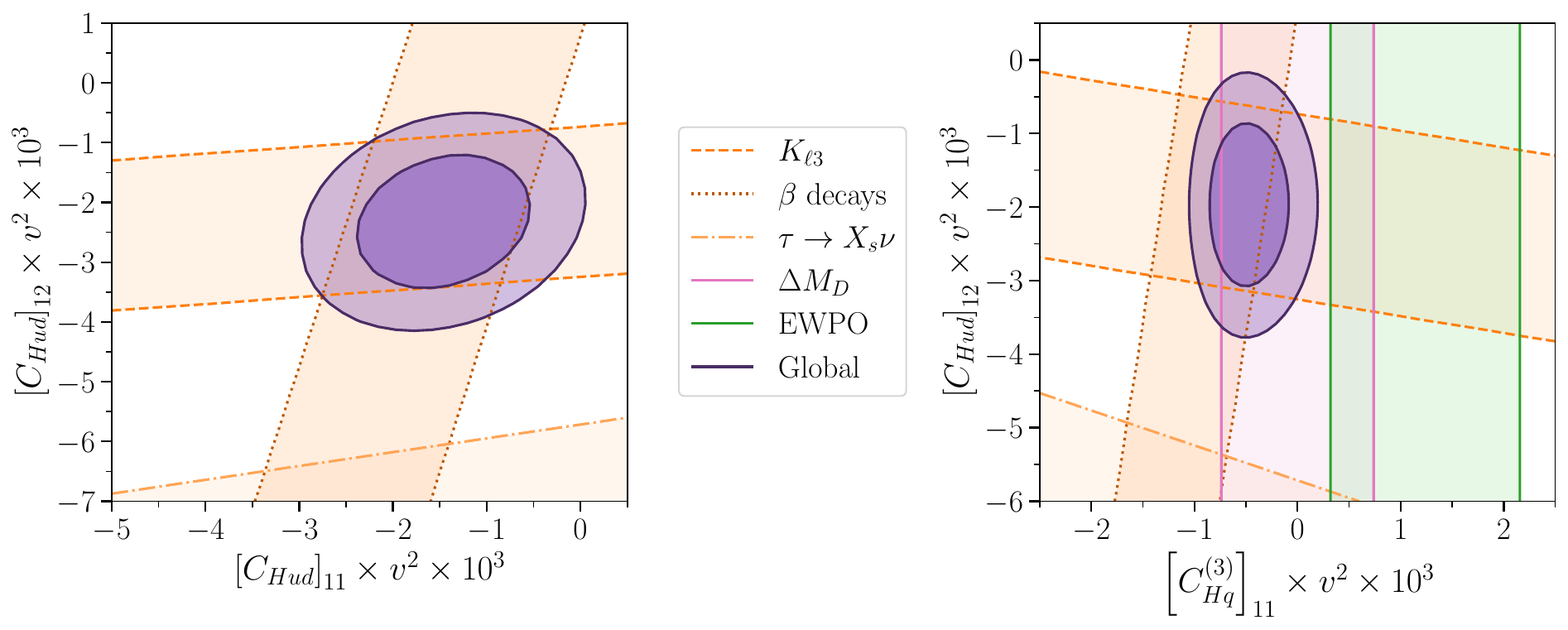}
\caption{Preferred regions for our two 2-D scenarios, see main text for details.}
\label{fig:eft_2d_combined}
\end{figure}

\paragraph{2-D scenarios}
As we have seen in the previous paragraph, non-zero values of $\CHudbracket_{11}$ and $\CHudbracket_{12}$ lead to modifications of right-handed $W$-$u$-$d$ and $W$-$u$-$s$ couplings and are able to solve and alleviate the tensions within $V_{us}$ and the unitarity deficit, respectively. The resulting preferred regions in the corresponding plane are shown in the left panel of \cref{fig:eft_2d_combined}. Note that while inclusive $\tau$ decays are not directly sensitive to right-handed currents, we get a constraint here since they modify the extraction of \V{us} (in our scheme), leading to an indirect sensitivity. On the other hand, despite exclusive $\tau$ decays being more precise, they do not present a constraint here as their theoretical prediction is affected in the same way as $K_{\mu2} / \pi_{\mu 2}$ used as an input in our scheme.
\smallskip

Alternatively we can consider non-zero values of $\CHqthreebracket_{11}$ and $\CHudbracket_{12}$ if we aim at explaining both tensions, leading to modifications of the left-handed $W$-$u$-$d$ and a right-handed $W$-$u$-$s$ couplings. The resulting preferred regions are shown on the right of \cref{fig:eft_2d_combined}. The best fit points of these two dimensional scenarios together with the pulls are given in Table~\ref{tab:eft_best_fits}.
\medskip

\paragraph{3-D scenario}

\begin{figure}
\includegraphics[width=0.6\textwidth]{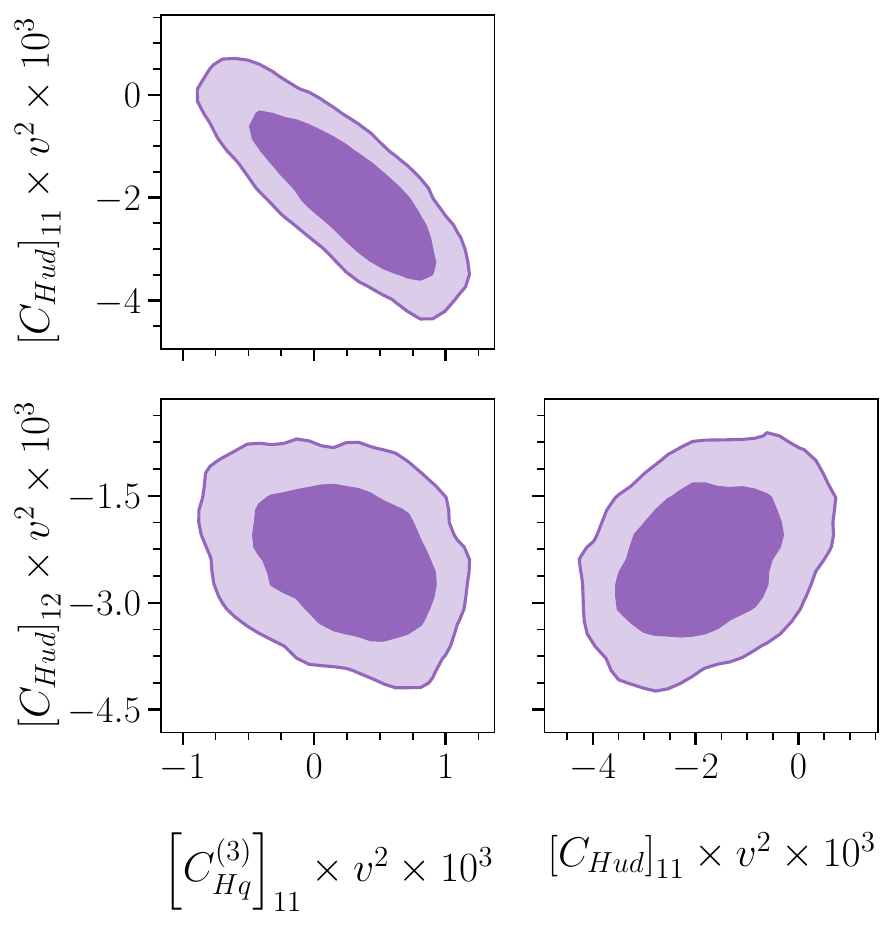}
\caption{Global fit to our 3-D scenario with non-zero Wilson coefficients $\CHqthreebracket_{11}$, $\CHudbracket_{11}$, and $\CHudbracket_{12}$.}
\label{fig:eft_LH_ud_RH_ud_us}
\end{figure}

Here we consider the modifications induced if $\CHqthreebracket_{11}$, $\CHudbracket_{11}$, and $\CHudbracket_{12}$ are simultaneously non-zero. 
The results are shown in \cref{fig:eft_LH_ud_RH_ud_us} from where we see that, similar to our 2-D scenarios, there is a strong preference for NP here, but also the significant correlation between left-handed and right-handed $W$-$u$-$d$ modifications. In the appendix~\ref{app:4d_scenarios}, we consider the 4-D scenarios in which we avoid or weaken the bounds from \Dz--\Dzb mixing by adding $\CHqthreebracket_{11}$ or a $\CHqthreebracket_{22}$ as free parameters. However, the situation does not change significantly compared to the 3-D scenario as can also be seen from the pulls given in Table~\ref{tab:eft_best_fits}.
\medskip

\paragraph{Summary}
The scenarios with modifications of right-handed $W$-$u$-$s$ couplings provide the best improvement relative to the SM (which roughly agrees with the results in Ref.~\cite{Grossman:2019bzp}) and do not lead to problems in flavour physics or electroweak precision measurements since constraints from $SU(2)_L$ invariance are not present. The scenarios with both left-handed and right-handed modifications displays a slightly larger $\Delta \chi^2$ (which can be understood by the fact left-handed operators change the EW fit by modifying $Z$-quark couplings) as summarized in \cref{tab:eft_best_fits}.

\begin{table}
\begin{tabular}{@{}LLLL@{}}
\toprule
\text{EFT Scenario} & \text{Best fit point} & - \Delta \chi^2 & \text{Pull} \\
\midrule 
\CHqthreebracket_{11} & -0.50 & 3.3 & \SI{1.8}{\sigma}
\\
\CHqthreebracket_{11} = \CHqthreebracket_{22} & -0.27 & 1.1 & \SI{1.1}{\sigma}
\\
\CHqthreebracket_{11} = \CHqonebracket_{11} & -0.55 & 3.7 & \SI{1.9}{\sigma}
\\
\CHudbracket_{11} & -1.0 & 3.1 & \SI{1.8}{\sigma}
\\
\CHudbracket_{12} & -2.0 & 7.4 & \SI{2.7}{\sigma}
\\
\left(\CHudbracket_{11}, \CHudbracket_{12}\right) & (-1.4, -2.1) & 13 & \SI{3.2}{\sigma} 
\\
\left(\CHqthreebracket_{11}, \CHudbracket_{12}\right) & (-0.43, -2.0) & 11 & \SI{2.8}{\sigma}
\\
\left(\CHqthreebracket_{11}, \CHudbracket_{11}, \CHudbracket_{12}\right) & (0.27, -1.9, -2.4) & 16 & \SI{2.9}{\sigma}
\\
\left(\CHqthreebracket_{11}, \CHqthreebracket_{22}, \CHudbracket_{11}, \CHudbracket_{12}\right) & (0.59, 0.76, -2.6, -2.5) & 17 & \SI{2.9}{\sigma}
\\
\left(\CHqthreebracket_{11}, \CHqonebracket_{11}, \CHudbracket_{11}, \CHudbracket_{12}\right) & (0.29, 0.11, -2.0, -2.4) & 13 & \SI{2.6}{\sigma}
\\
\bottomrule
\end{tabular}
\caption{Best fit points, $\Delta \chi^2$ and pulls w.r.t.~the SM hypothesis for the various EFT scenarios. The best fit points are in units of $10^{-3} v^{-2}$.
\label{tab:eft_best_fits}}
\end{table}

\subsection{Vector-like Quark models}
\label{sec:vlqs}

Now we examine VLQs coupling to first and second generation quarks in general, and the representations providing a potential solution to the tensions in the CKM matrix in particular.\footnote{For a recent analysis of VLQs coupling to third generation quarks we refer the interested reader to Ref.~\cite{Crivellin:2022fdf}.}
We fix the masses of the VLQs to \SI{2}{\TeV}, which is compatible with LHC searches~\cite{ATLAS:2011tvb,ATLAS:2012apa,ATLAS:2015lpr,CMS:2017asf} (a recent study has shown that the high-luminosity LHC could exclude a first generation $U$ VLQ at this mass for $\xi^U_1 \gtrsim 0.25$~\cite{Cui:2022hjg}).
Note that the scaling of the bounds (with the exception of $\Delta F=2$ processes) is just proportional to coupling squared over mass squared, modulus logarithmic effects from the renormalization group evolution. For \Dz--\Dzb and kaon mixing, we have included the one-loop matching which becomes relevant for larger masses and breaks the simple scaling observed in the other processes.
\smallskip

\begin{figure}
\includegraphics[width=\textwidth]{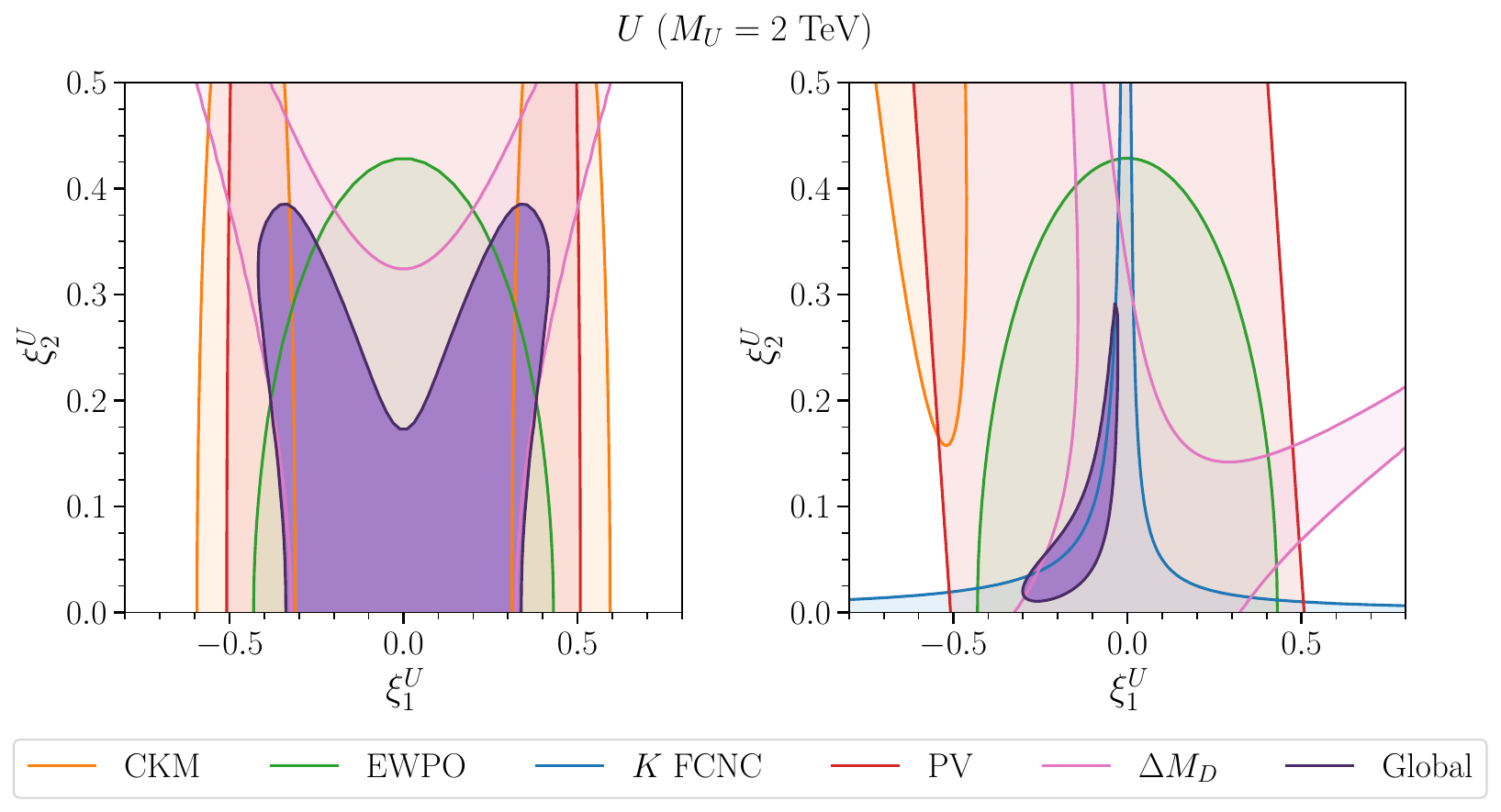}
\caption{Global fit to $1^{\rm st}$ and $2^{\rm st}$ generation couplings of the VLQ $U$. The left-hand side assumes multiple generations of VLQs, each only coupling to a single generation (i.e.\ effectively removing  constraints from kaon physics), while on the right side this assumption is removed.}
\label{fig:2d_U_fit}
\end{figure}

\begin{figure}
\includegraphics[width=\textwidth]{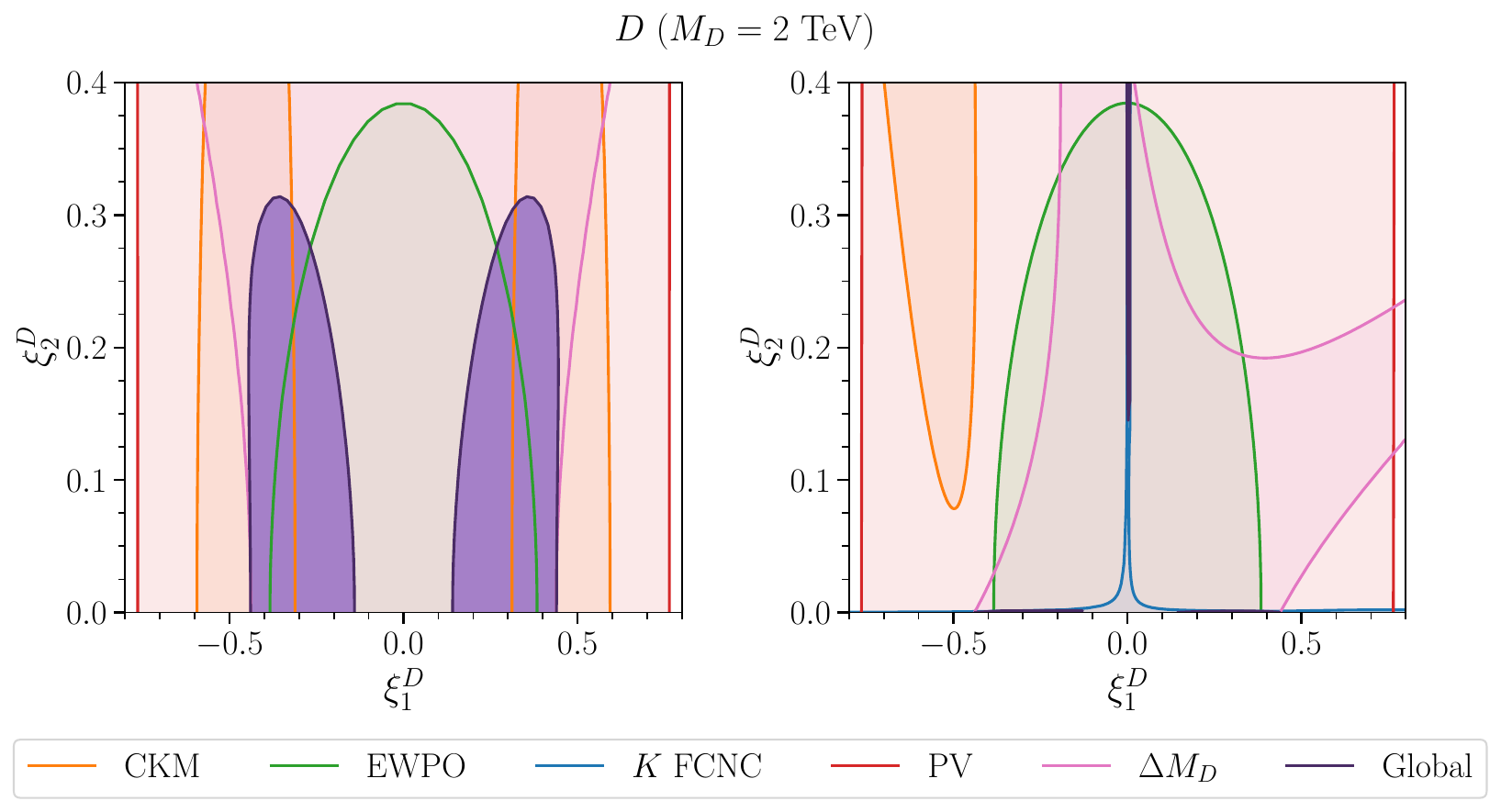}
\caption{Global fit to $1^{\rm st}$ and $2^{\rm st}$ generation couplings of the VLQ $D$. The left-hand side assumes multiple generations of VLQs, each only coupling to a single generation (i.e.\ effectively removing  constraints from kaon physics), while on the right side this assumption is removed.}
\label{fig:2d_D_fit}
\end{figure}

\begin{figure}
\includegraphics[width=\textwidth]{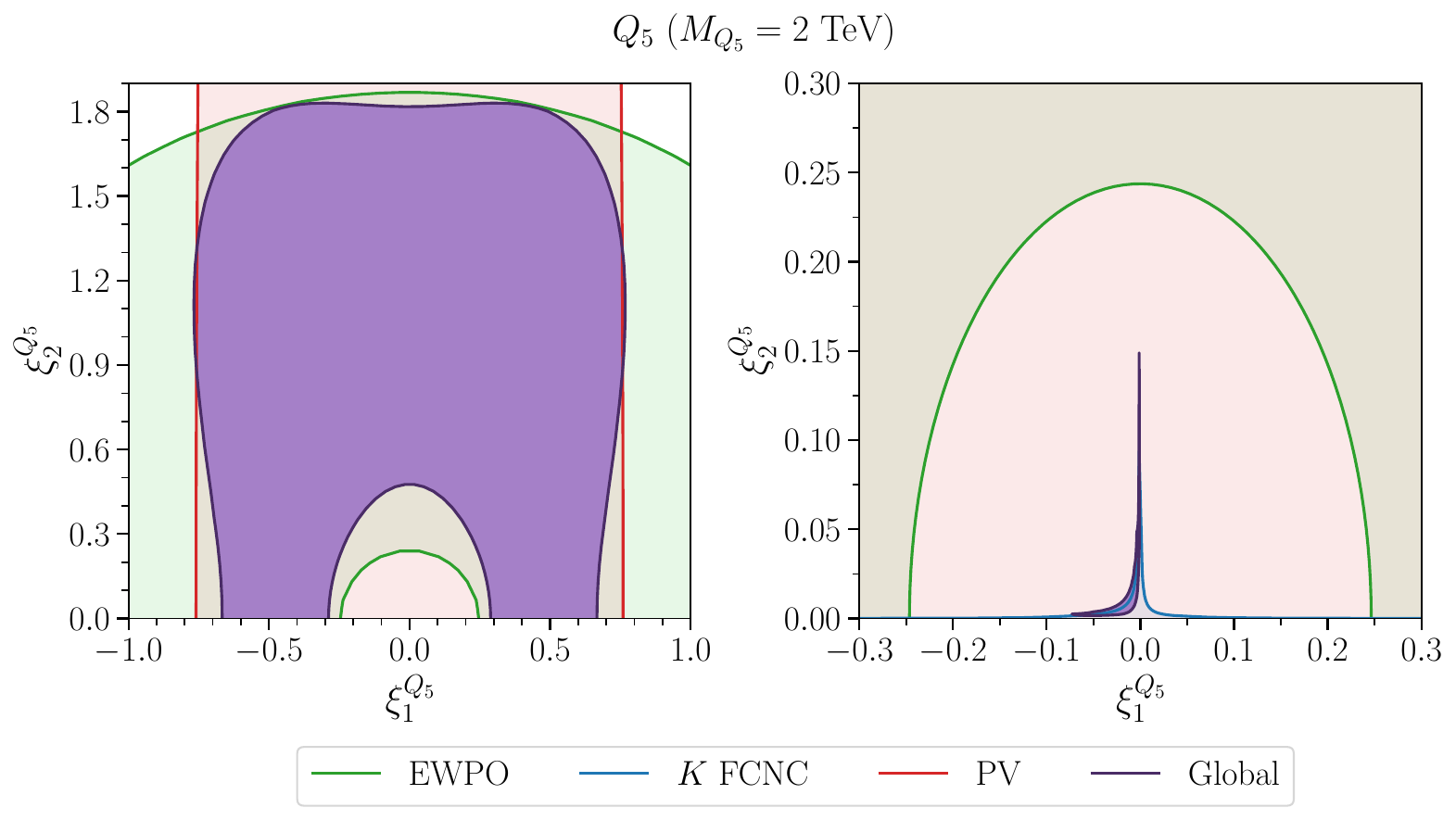}
\caption{Global fit to $1^{\rm st}$ and $2^{\rm st}$ generation couplings of the VLQ $Q_5$. The left-hand side assumes multiple generations of VLQs, each only coupling to a single generation (i.e.\ effectively removing  constraints from kaon physics), while on the right side this assumption is removed.}
\label{fig:2d_Q5_fit}
\end{figure}

\begin{figure}
\includegraphics[width=\textwidth]{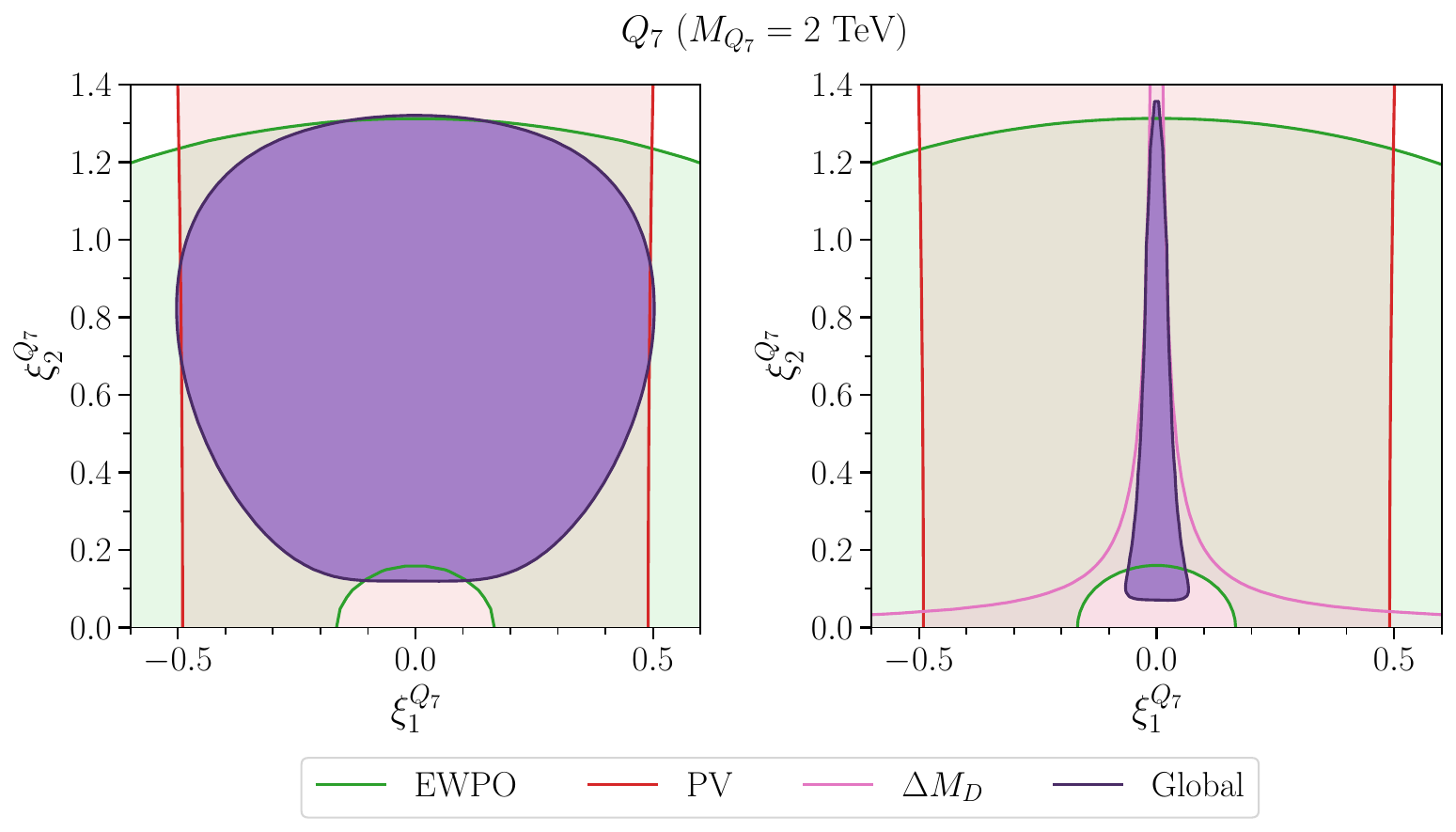}
\caption{Global fit to $1^{\rm st}$ and $2^{\rm st}$ generation couplings of the VLQ $Q_7$. The left-hand side assumes multiple generations of VLQs, each only coupling to a single generation (i.e.\ effectively removing  constraints from kaon physics), while on the right side this assumption is removed.}
\label{fig:2d_Q7_fit}
\end{figure}

\begin{figure}
\includegraphics[width=\textwidth]{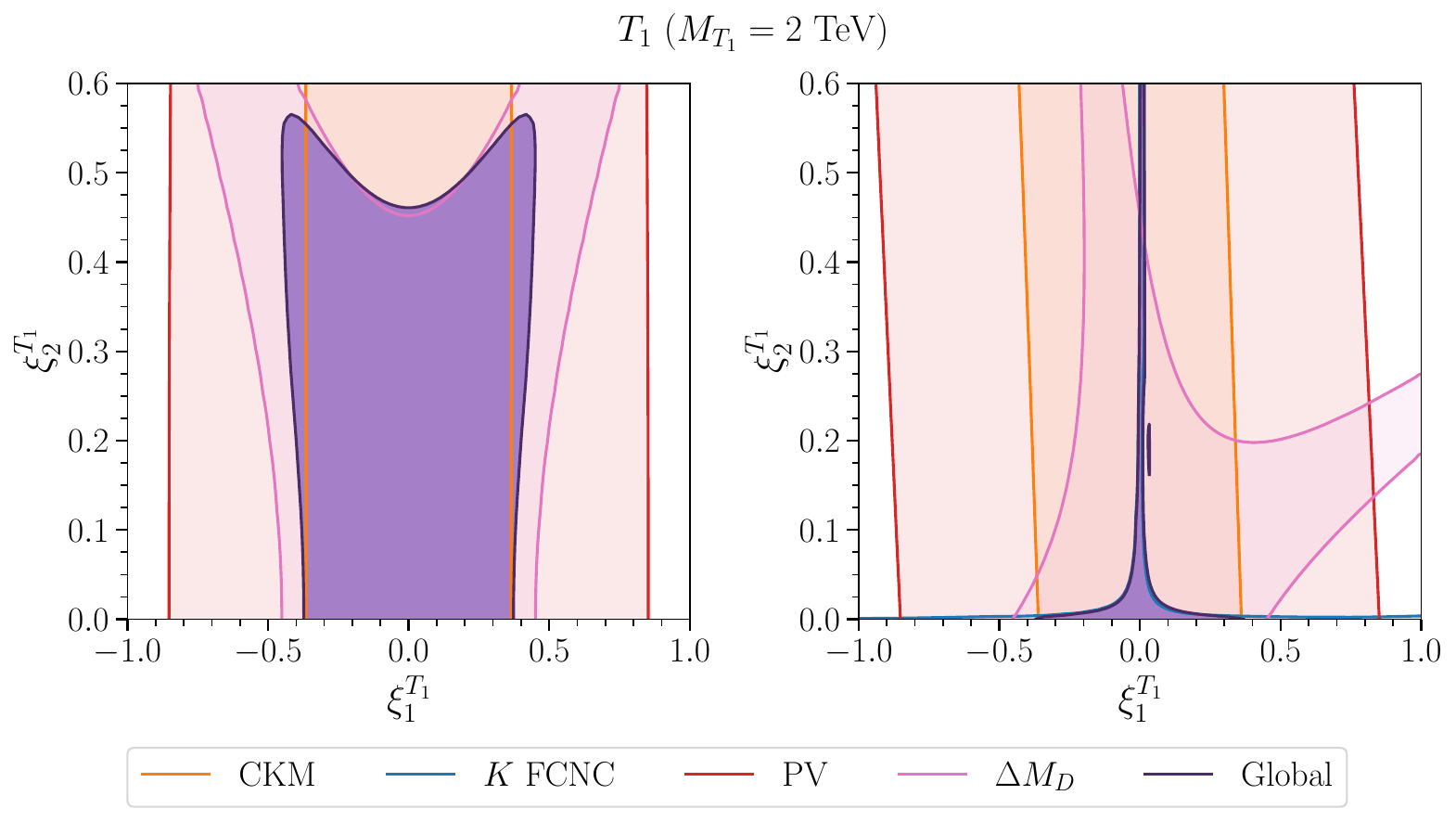}
\caption{Global fit to $1^{\rm st}$ and $2^{\rm st}$ generation couplings of the VLQ $T_1$. The left-hand side assumes multiple generations of VLQs, each only coupling to a single generation (i.e.\ effectively removing  constraints from kaon physics), while on the right side this assumption is removed.}
\label{fig:2d_T1_fit}
\end{figure}

\begin{figure}
\includegraphics[width=\textwidth]{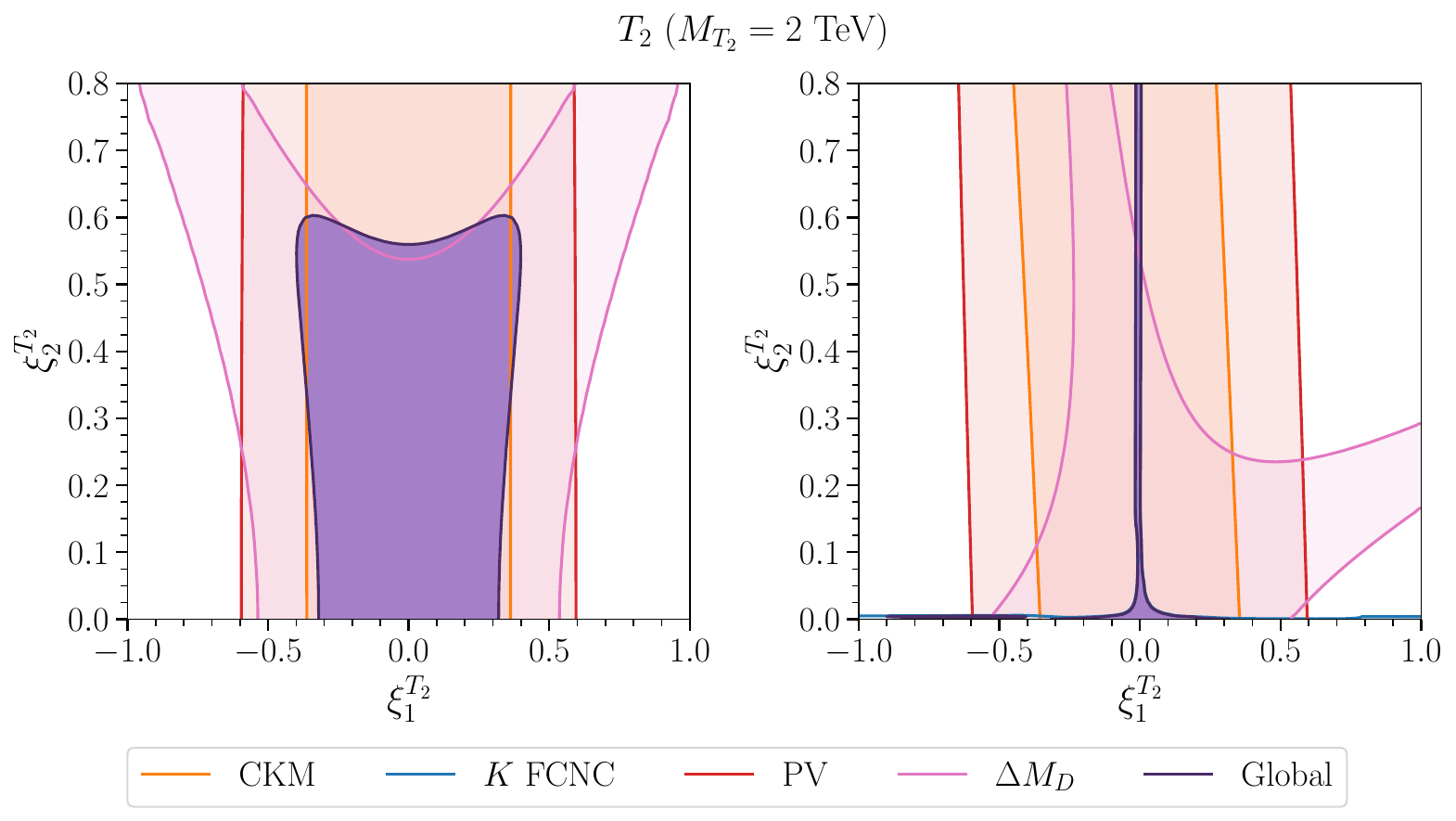}
\caption{Global fit to $1^{\rm st}$ and $2^{\rm st}$ generation couplings of the VLQ $T_2$. The left-hand side assumes multiple generations of VLQs, each only coupling to a single generation (i.e.\ effectively removing  constraints from kaon physics), while on the right side this assumption is removed.}
\label{fig:2d_T2_fit}
\end{figure}

In the \cref{fig:2d_U_fit,fig:2d_D_fit,fig:2d_Q5_fit,fig:2d_Q7_fit,fig:2d_T1_fit,fig:2d_T2_fit} we show in the left-handed panels the preferred regions assuming multiple generations of VLQs, coupling separately to first and second generations quarks, thus avoiding tree-level effects in kaon FCNC processes (despite effects from CKM rotations in \Dz--\Dzb mixing). In the right-handed panels the same fit for a single LQ representations, coupling simultaneously to first and second generations quarks is shown. Here, as explained in Sec.~\ref{sef:fitmethod}, ``CKM'' stands for the combined region from the observables listed in \cref{tab:smelli_Kl3_obs,tab:smelli_beta+n_obs,tab:smelli_tau_excl_obs,tab:smelli_Ddecay_obs} as well as from inclusive $\tau$ decay, while the ``$K$ FCNC'' includes the observables listed in \cref{tab:smelli_K_FCNC_obs}. Let us now discuss the various representations separately.

\paragraph{$U$ (\cref{fig:2d_U_fit})}
The $SU(2)_L$ singlet $U$ (with quantum numbers of a right-handed up-type quark of the SM) leads to modified left-handed $W$ coupling to quarks, so that the CKM tensions favour a non-zero first generation coupling. However, EW precision measurements and data from PV experiments limit the possible size of this coupling, even in the absence of direct contributions to \Dz--\Dzb mixing (left panel).
In the right panel, the best fit point is at $\xi^U_1 = -0.2, \xi^U_2 = 0.045$ with a pull w.r.t.\ the SM of \SI{2.2}{\sigma}.
\smallskip

\paragraph{$D$ (\cref{fig:2d_D_fit})}

The allowed regions for the Yukawa couplings of $D$ (which has the quantum numbers of a right-handed down-type quark in the SM). We see that while for a single generation kaon FCNC constraints are very severe, (right panel) while in the situation with two generations the allowed regions are much more sizable (left panel).
We see that in either case, the data favours a single non-zero coupling, with a best fit at $\xi^D_1 = -0.34$ and a one-dimensional pull w.r.t.\ the SM of \SI{1.8}{\sigma}.
 \smallskip

\paragraph{$Q_5$ (\cref{fig:2d_Q5_fit})}
This $SU(2)_L$ doublet VLQ with exotic hypercharge only generates modified $Z$ couplings (but no $W$ couplings) at tree-level, and hence there is no sensitivity to the CKM anomalies. In fact, as the modifications are only to the right-handed $Z$-$d$-$d$ couplings, the current bounds on this VLQ are very weak, as can be seen in the left-hand side of the figure.
While PV provides some bounds, there is a small preference for non-zero couplings from EW precision measurements due to the current small tensions in $Z$ width and hadronic cross-section results. Once we allow for a single VLQ to couple to both generations however, we find that kaon physics drastically reduces the allowed region -- this occurs due to the RG and matrix element enhancement of $(\bar s \gamma^\mu P_L d)(\bar s \gamma_\mu P_R d)$ four-quark operators in $\epsilon_K$.
\medskip

\paragraph{$Q_7$ (\cref{fig:2d_Q7_fit})}
The results for this VLQ are very similar to the ones for $Q_5$, even though this VLQ modifies right-handed $Z$-$u$-$u$ couplings, instead of $Z$-$d$-$d$ ones, although now the main bounds, in case it couples to both first and second generation, originate from \Dz--\Dzb mixing.
\medskip

\paragraph{$T_1$ (\cref{fig:2d_T1_fit})}
This $SU(2)_L$ triplet modifies the left-handed charged current, thus affecting the CKM determinations, but in the wrong direction to resolve the first-row unitarity deviations. Therefore, the CKM measurements merely provide a constraint on its interactions, alongside \Dz--\Dzb mixing and parity violation. The modifications to $Z$-quark couplings are smaller than in case of the $SU(2)_L$ singlet VLQs, and so the corresponding bounds cannot be seen in our region shown. Once we allow the triplet to couple to both generations at once, \Dz--\Dzb mixing becomes stronger and kaon constraints are extremely tight.
\medskip

\paragraph{$T_2$ (\cref{fig:2d_T2_fit})}
The other triplet $T_2$ has essentially the same bounds as the first, but bounds from low-energy parity violation are slightly stronger while \Dz--\Dzb mixing slightly weaker. Again, once we include couplings to both generations of a single VLQ, kaon decays prove to be very strong and the globally allowed region is quite small.
\medskip

\begin{figure}
\includegraphics[width=\textwidth]{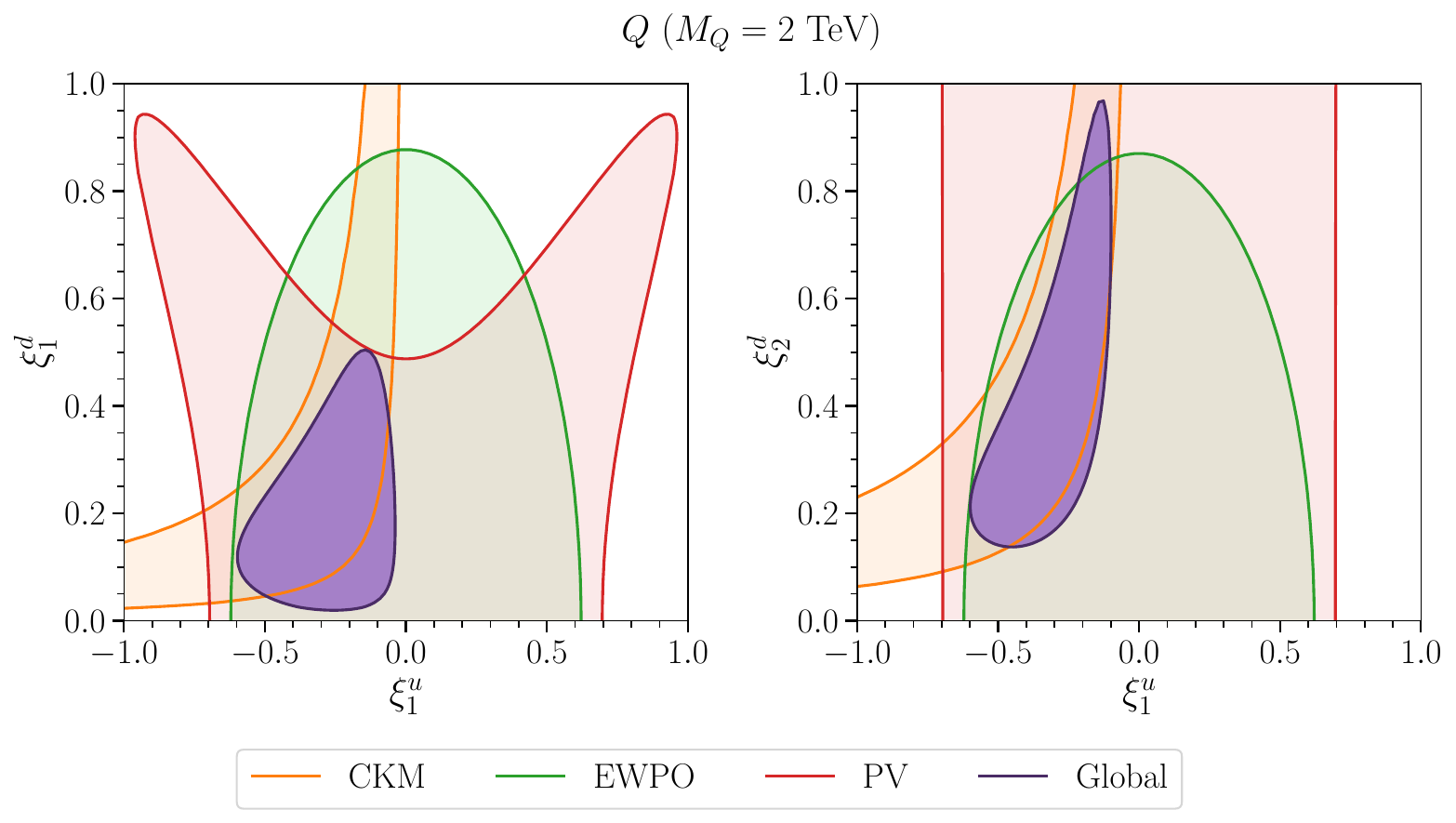}
\caption{Preferred regions for the VLQ $Q$ with either couplings to $u$ and $d$ (left) or $u$ and $s$ (right).}
\label{fig:2d_Q1_fits}
\end{figure}

\paragraph{$Q$ (\cref{fig:2d_Q1_fits})}

For the $SU(2)_L$ doublet $Q$, since it can have both couplings to right-handed up and down quarks, we instead show two fits, for either purely $1^{\rm st}$ or purely $2^{\rm nd}$ generation down quark interactions. The $Q$ doublet is unique in generating modifications to the right-handed $W$ couplings and, as expected from our previous EFT results, there is a strong preference towards non-zero couplings to both right-handed $u$ and $d$ (left) or $u$ and $s$ (right).
However, unlike in our simple EFT scenario, the $Q$ field generates additional correlated effects in $Z$ couplings through $SU(2)$ invariance, and so PV and EWPO partially limit the parameter space. 
In the left panel, the best fit point is $\xi^u_1 = -0.29$, $\xi^d_1 = 0.21$ and has a pull w.r.t.\ the SM of \SI{1.1}{\sigma}. For the right panel, we find a best fit at $\xi^u_1 = -0.33$, $\xi^d_2 = 0.38$ and a pull of \SI{2.1}{\sigma}.
\medskip

\paragraph{VLQs and tensions in the Cabibbo angle}

In our SMEFT analysis we saw that resolving the CAA via modified gauge boson couplings requires NP in $\CHqthreebracket_{11}$ and/or $\CHudbracket_{11}$ as well as in $\CHudbracket_{12}$ for the $V_{us}$ tension. From the tree-level matching of VLQs on the SMEFT (see Eq.~\eqref{eq:VLQ_SMEFT_tree_matching}), we see that only the $SU(2)_L$ doublet $Q$ can generate the coefficients $\CHud$ that generate right-handed $W$ couplings. Furthermore, for a single generation of the doublet, NP in the right-handed $W$-$u$-$d$ and $W$-$u$-$s$ vertices (at the same time) lead to significant NP in right-handed $Z$-$d$-$s$ couplings, stringently constrained by $\epsilon_K$~\cite{Endo:2016tnu,Bobeth:2017xry}. Updating their result, we find that a single $Q$ doublet coupled to both $d$ and $s$ would have to obey $M_Q / \sqrt{\xi^d_1 \xi^d_2} > \SI{175}{\TeV}$ to be consistent with experiment, and therefore far too heavy to be relevant to the CAA.
\smallskip

Thus, a full explanation of the tensions in the Cabibbo angle determination require a modified $W$-$u$-$d$ and $W$-$u$-$s$ coupling and thus multiple generations of $Q$. Similarly, one can solve the CAA via a modified left-handed $W$-$u$-$d$ coupling and a right-handed $W$-$u$-$s$ coupling which again requires at least two VLQs. This means that a full solution of the CAA demands the presence of at least two VLQs.
\medskip

\section{Conclusions}
\label{sec:conclu}

In this article we studied modified couplings of light quarks to EW gauge bosons, both in the SMEFT and in models with VLQs. We paid particular attention to the different determinations of the Cabibbo angle that are in disagreement with each other, pointing towards such modified $W$ couplings to quarks. We performed a global analysis using \smelli, taking into account the constraints for EW precision and flavour observables.
\smallskip

In more detail, we first summarised the current status of the determinations of the CKM elements \V{ud} and \V{us}. For \V{ud}, where super-allowed $\beta$ decays continue to provide the most precise determination, neutron decays is quickly approaching competitively, and its current central value is only slightly larger than (and therefore perfectly consistent with) the one from super-allowed decays. For the direct determination of \V{us} the leading decay mode is $K \to \pi \ell \nu$ as there are still questions to be resolved regarding theory prediction for inclusive $\tau$ decays and how this should be applied to data. Finally, the ratio \V{us}/\V{ud} is dominated by $K \to \mu \nu / \pi \to \mu \nu$ as the ratios $\tau \to K \nu / \tau \to \pi \nu$ and $K \to \pi \ell \nu / \pi \to \pi e \nu$ are currently limited by the experimental data. We tested the CKM unitarity prediction of the SM by taking each pair of the best measurements individually, and find that it is violated between \SI{1.8}{\sigma} and \SI{3.3}{\sigma} (see Eq.~\eqref{eq:testunitarity}). This agrees with the result of the global fit to all available data where we find the unitarity violation at the \SI{2.8}{\sigma} level (see Eq.~\eqref{eq:CAAglobal}). Furthermore, a tension between the different determinations of \V{us} exists, which can only be explained by non-standard right-handed interactions and can be tested  through a $K \to \pi \mu \nu/ K \to \mu \nu$ measurement in the near future
 by NA62~\cite{Cirigliano:2022yyo}.
\smallskip

In our global SMEFT fit, we found that several scenarios can solve or alleviate the tensions in the determinations of the Cabibbo angle, as summarized in \cref{tab:eft_best_fits}. The simplest and least problematic case is that of right-handed charged currents (both in $W$-$u$-$d$ and $W$-$u$-$s$ couplings) which can bring all the main determinations into agreement without being in conflict with EW precision or flavour observables, and is therefore favoured over the SM hypothesis by \SI{3.2}{\sigma} (\SI{2.7}{\sigma}) in the best 2-D (1-D) scenario. Other scenarios with modified left-handed charged currents (only) are also preferred over the SM hypothesis, but cannot account for the discrepancies within $V_{us}$ and face constraints from EW precision measurements as well as \Dz--\Dzb mixing. While the latter bounds can be weakened or avoided by considering a $U(2)$ flavour symmetry or a specific combination of Wilson coefficients, respectively, the global fit displays a maximal pull of $1.9\sigma$ if only left-handed currents are considered. 
\smallskip

The most natural extension of the SM that leads to modified EW gauge couplings to quarks are VLQs. They affect these couplings already at tree-level and are also theoretically well-motivated, e.g.,~by grand unified theories, composite and extra-dimensional models and little Higgs models etc.
We first matched the different VLQ representations under the SM gauge group on the SMEFT (at tree-level for the charged current and EW precision observables and at loop-level for $\Delta F=2$ processes) and used these results to calculate the relevant effects in the related observables. We then performed a global fit for the different representations of VLQs, taking into account couplings to first and second generation quarks. While a single VLQ coupling simultaneously to first and second generation quarks will lead to FCNCs and is hence very constrained, these bounds can be avoided or weakened for multiple generations of VLQs.
\smallskip

The $SU(2)_L$ singlet VLQs can improve the fit w.r.t.~the CAA, but EW precision and \Dz--\Dzb mixing (as well as PV measurements for the $U$)  prevent a better description of data. The $SU(2)_L$ triplet VLQs generate the wrong sign to match our left-handed EFT scenario for the CAA, so that here CKM unitarity acts as a constraint but the tension cannot be explained. The $SU(2)_L$ doublets with non-SM-like hypercharges do not contribute to CKM observables, as they only generate modified $Z$ but not $W$ couplings to quarks. The heavy SM-like doublet $Q$ proves the most interesting case, as this is the only VLQ that generates the right-handed $W$ couplings to quarks. As expected from our EFT fits, this VLQ is strongly favoured by the CKM measurements, but now faces bounds from EWPO and PV as modified right-handed $Z$ couplings to quarks are also induced, removing some of the parameter space and reducing the improvement to the $\chi^2$ of the fit to data.
In fact, if one compares the $\CHudbracket_{11}$ scenario with a best fit pull of \SI{1.8}{\sigma} to the corresponding $Q$ UV model with only \SI{1.1}{\sigma}, and similarly the $\CHudbracket_{12}$ scenario has a best fit pull of \SI{2.7}{\sigma}, compared to only \SI{2.1}{\sigma} in the second $Q$ scenario.
We note however the best fit points remains consistent with the right-handed EFT fit.
\smallskip

In conclusion, the tensions related to the determination of the Cabibbo angle can be most easily explained by new physics leading to right-handed charged currents and therefore by vector-like quark $Q$. Therefore, while collider bounds for third generations VLQs have been well studied, the CAA provides strong motivation for searches for VLQs coupling to first and second generation quarks.
\medskip


\acknowledgments{
We would like to thank Martin Hoferichter and Alberto Lusiani for useful discussions.
We would also like to thank Eugenio Nicotina for noticing a sign error in \cref{eq:WZCH} in an earlier version of this article.
A.C.~acknowledges financial support by the Swiss National Science Foundation, Project No. PP00P\_2176884. T.K.~is supported by the Grant-in-Aid for Early-Career Scientists (No.\,19K14706) and by the JSPS Core-to-Core  Program (Grant No.\,JPJSCCA20200002) from the Ministry of Education, Culture, Sports, Science, and Technology (MEXT), Japan.
M.K.\ and F.M.\ acknowledge support from a Maria Zambrano fellowship, and from the State Agency for Research of the Spanish Ministry of Science and Innovation through the “Unit of Excellence Mar\'ia de Maeztu 2020-2023” award to the Institute of Cosmos Sciences (CEX2019-000918-M) and from PID2019-105614GB-C21 and 2017-SGR-929 grants. M.K.\ also acknowledges previous support by MIUR (Italy) under a contract PRIN 2015P5SBHT and by INFN Sezione di Roma La Sapienza, by the ERC-2010 DaMESyFla Grant Agreement Number: 267985. A.C.\ and M.K.\ would like to thank the Aspen Center for Physics (National Science Foundation grant PHY-1607611) and the Mainz Institute for Theoretical Physics (MITP) of the Cluster of Excellence PRISMA+ (Project ID 39083149) for their hospitality and support.
M.K.\ was supported by a grant from the Simons Foundation while at Aspen.
}


\appendix

\section{SMEFT matching}
\label{app:smeft_matching}

Here we present the matching expressions at one-loop for the four-quarks operators which contribute to \Dz--\Dzb mixing, which we calculated using \mmeft \cite{Carmona:2021xtq}.\footnote{A partial calculation of the one-loop matching was done in \cite{Bobeth:2016llm} for the $D$, $Q$, $Q_5$, $T_1$, and $T_2$ VLQs.}

\begin{align}
\left[ C_{qq}^{(1)} \right]_{ijkl} &= 
    \begin{aligned}[t]
   &- \frac{\xi^{U *}_{fi} \xi^U_{fj} \xi^{U *}_{gk} \xi^{U}_{gl}}{256 \pi^2 M_{U}^2}
    + \frac{3 (\Yukawa{u} \Yukawadagger{u})_{ij} \xi^{U *}_{fk} \xi^{U}_{fl}}{512 \pi^2 M_{U}^2}
    + \frac{3 (\Yukawa{u} \Yukawadagger{u})_{kl} \xi^{U *}_{fi} \xi^{U}_{fj}}{512 \pi^2 M_{U}^2}
    \\
   &- \frac{\xi^{D *}_{fi} \xi^D_{fj} \xi^{D *}_{gk} \xi^{D}_{gl}}{256 \pi^2 M_{D}^2}
    - \frac{3 (\Yukawa{u} \Yukawadagger{u})_{ij} \xi^{D *}_{fk} \xi^{D}_{fl}}{512 \pi^2 M_{D}^2}
    - \frac{3 (\Yukawa{u} \Yukawadagger{u})_{kl} \xi^{D *}_{fi} \xi^{D}_{fj}}{512 \pi^2 M_{D}^2}
    \\
   &- \frac{9 \xi^{T_1 *}_i \xi^{T_2}_j \xi^{T_1 *}_k \xi^{T_2}_l}{4096 \pi^2 M_{T_1}^2}
    - \frac{9 (\Yukawa{u} \Yukawadagger{u})_{ij} \xi^{T_1 *}_{fk} \xi^{T_1}_{fl}}{2048 \pi^2 M_{T_1}^2}
    - \frac{9 (\Yukawa{u} \Yukawadagger{u})_{kl} \xi^{T_1 *}_{fi} \xi^{T_1}_{fj}}{2048 \pi^2 M_{T_1}^2}
    \\
   &- \frac{9 \xi^{T_2 *}_{fi} \xi^{T_2}_{fj} \xi^{T_2 *}_{gk} \xi^{T_2}_{gl}}{4096 \pi^2 M_{T_2}^2}
    + \frac{9 (\Yukawa{u} \Yukawadagger{u})_{ij} \xi^{T_2 *}_{fk} \xi^{T_2}_{fl}}{2048 \pi^2 M_{T_2}^2}
    + \frac{9 (\Yukawa{u} \Yukawadagger{u})_{kl} \xi^{T_2 *}_{fi} \xi^{T_2}_{fj}}{2048 \pi^2 M_{T_2}^2}
     \,,
    \end{aligned}
\\
\left[ C_{qq}^{(3)} \right]_{ijkl} &= 
    \begin{aligned}[t]
   &- \frac{\xi^{U *}_{fi} \xi^U_{fj} \xi^{U *}_{gk} \xi^{U}_{gl}}{256 \pi^2 M_{U}^2}
    + \frac{3 (\Yukawa{u} \Yukawadagger{u})_{ij} \xi^{U *}_{fk} \xi^{U}_{fl}}{512 \pi^2 M_{U}^2}
    + \frac{3 (\Yukawa{u} \Yukawadagger{u})_{kl} \xi^{U *}_{fi} \xi^{U}_{fj}}{512 \pi^2 M_{U}^2}
    \\
   &- \frac{\xi^{D *}_{fi} \xi^D_{fj} \xi^{D *}_{gk} \xi^{D}_{gl}}{256 \pi^2 M_{D}^2}
    + \frac{3 (\Yukawa{u} \Yukawadagger{u})_{ij} \xi^{D *}_{fk} \xi^{D}_{fl}}{512 \pi^2 M_{D}^2}
    + \frac{3 (\Yukawa{u} \Yukawadagger{u})_{kl} \xi^{D *}_{fi} \xi^{D}_{fj}}{512 \pi^2 M_{D}^2}
    \\
   &- \frac{ \xi^{T_1 *}_{fi} \xi^{T_2}_{fj} \xi^{T_1 *}_{gk} \xi^{T_2}_{gl}}{4096 \pi^2 M_{T_1}^2}
    - \frac{3 (\Yukawa{u} \Yukawadagger{u})_{ij} \xi^{T_1 *}_{fk} \xi^{T_1}_{fl}}{2048 \pi^2 M_{T_1}^2}
    - \frac{3 (\Yukawa{u} \Yukawadagger{u})_{kl} \xi^{T_1 *}_{fi} \xi^{T_1}_{fj}}{2048 \pi^2 M_{T_1}^2}
    \\
   &- \frac{\xi^{T_2 *}_{fi} \xi^{T_2}_{fj} \xi^{T_2 *}_{gk} \xi^{T_2}_{gl}}{4096 \pi^2 M_{T_2}^2}
    - \frac{3 (\Yukawa{u} \Yukawadagger{u})_{ij} \xi^{T_2 *}_{fk} \xi^{T_2}_{fl}}{2048 \pi^2 M_{T_2}^2}
    - \frac{3 (\Yukawa{u} \Yukawadagger{u})_{kl} \xi^{T_2 *}_{fi} \xi^{T_2}_{fj}}{2048 \pi^2 M_{T_2}^2}
     \,,
    \end{aligned}
\\
\left[ C_{qu}^{(1)} \right]_{ijkl} &= 
    \begin{aligned}[t]
   &- \frac{3 (\Yukawadagger{u} \Yukawa{u})_{kl} \xi^{U *}_{fi} \xi^{U}_{fj}}{128 \pi^2 M_{U}^2}
    - \frac{\Yukawa{u}_{il} \Yukawadagger{u}_{kj} \xi^{U *}_{fm} \xi^{U}_{fm}}{96 \pi^2 M_{U}^2}
+ \frac{3 (\Yukawadagger{u} \Yukawa{u})_{kl} \xi^{D *}_{fi} \xi^{D}_{fj}}{128 \pi^2 M_{D}^2}    \\
   &
    + \frac{\Yukawa{u}_{il} \xi^D_{fj} (\Yukawadagger{u} \xi^{D \dagger})_{kf}}{192 \pi^2 M_D^2}
    + \frac{\xi^{D *}_{fi} \Yukawadagger{u}_{kj} (\xi^D \Yukawa{u})_{fl}}{192 \pi^2 M_D^2}
- \frac{\Yukawa{u}_{il} \Yukawadagger{u}_{kj} \xi^{D *}_{fm} \xi^{D}_{fm}}{96 \pi^2 M_{D}^2}
    \\
   &- \frac{3 (\Yukawa{u} \Yukawadagger{u})_{ij} \xi^{u *}_{fk} \xi^{u}_{fl}}{128 \pi^2 M_{Q}^2}
    - \frac{\Yukawa{u}_{il} \Yukawadagger{u}_{jk} \xi^{d *}_{fm} \xi^{d}_{fm}}{96 \pi^2 M_{Q}^2}
    - \frac{\Yukawa{u}_{il} \Yukawadagger{u}_{jk} \xi^{u *}_{fm} \xi^{u}_{fm}}{96 \pi^2 M_{Q}^2}
- \frac{\xi^{Q_5}_{fm} \xi^{Q_5 *}_{fm} \Yukawa{u}_{il} \Yukawadagger{u}_{kj}}{96 \pi^2 M_{Q_5}^2}
    \\
   &+ \frac{3 (\Yukawa{u} \Yukawadagger{u})_{ij} \xi^{Q_7 *}_{fk} \xi^{Q_7}_{fl}}{128 \pi^2 M_{Q_7}^2}
    - \frac{\Yukawa{u}_{il} (\xi^{Q_7} \Yukawadagger{u})_{fj} \xi^{Q_7 *}_{fk}}{192 \pi^2 M_{Q_7}^2}
    - \frac{(\Yukawa{u} \xi^{Q_7 \dagger})_{fi} \Yukawadagger{u})_{kj} \xi^{Q_7}_{fl}}{192 \pi^2 M_{Q_7}^2}
    \\ &- \frac{\xi^{Q_7}_{fm} \xi^{Q_7 *}_{fm} \Yukawa{u}_{il} \Yukawadagger{u}_{kj}}{96 \pi^2 M_{Q_7}^2}
    \\
   &+ \frac{9 (\Yukawadagger{u} \Yukawa{u})_{kl} \xi^{T_1 *}_{fi} \xi^{T_1}_{fj}}{512 \pi^2 M_{T_1}^2}
    - \frac{\Yukawa{u}_{il} \xi^D_{fj} (\Yukawadagger{u} \xi^{T_1 \dagger})_{kf}}{256 \pi^2 M_{T_1}^2}
    - \frac{\xi^{T_1 *}_{fi} \Yukawadagger{u}_{kj} (\xi^{T_1} \Yukawa{u})_{fl}}{256 \pi^2 M_{T_1}^2}
    \\ &- \frac{\Yukawa{u}_{il} \Yukawadagger{u}_{kj} \xi^{T_1 *}_{fm} \xi^{T_1}_{fm}}{128 \pi^2 M_{T_1}^2}
- \frac{9 (\Yukawadagger{u} \Yukawa{u})_{kl} \xi^{T_2 *}_{fi} \xi^{T_2}_{fj}}{512 \pi^2 M_{T_2}^2}
    - \frac{\Yukawa{u}_{il} \Yukawadagger{u}_{kj} \xi^{T_2 *}_{fm} \xi^{T_2}_{fm}}{128 \pi^2 M_{T_2}^2}
     \,,
    \end{aligned}
\\
\left[ C_{qu}^{(8)} \right]_{ijkl} &= 
    \begin{aligned}[t]
   &- \frac{\Yukawa{u}_{il} \Yukawadagger{u}_{kj} \xi^{U *}_{fm} \xi^{U}_{fm}}{16 \pi^2 M_{U}^2}
    \\
   &+ \frac{\Yukawa{u}_{il} \xi^D_{fj} (\Yukawadagger{u} \xi^{D \dagger})_{kf}}{32 \pi^2 M_D^2}
    + \frac{\xi^{D *}_{fi} \Yukawadagger{u}_{kj} (\xi^D \Yukawa{u})_{fl}}{32 \pi^2 M_D^2}
    - \frac{\Yukawa{u}_{il} \Yukawadagger{u}_{kj} \xi^{D *}_{fm} \xi^{D}_{fm}}{16 \pi^2 M_{D}^2}
    \\
   &- \frac{\Yukawa{u}_{il} \Yukawadagger{u}_{jk} \xi^{d *}_{fm} \xi^{d}_{fm}}{16 \pi^2 M_{Q}^2}
    - \frac{\Yukawa{u}_{il} \Yukawadagger{u}_{jk} \xi^{u *}_{fm} \xi^{u}_{fm}}{16 \pi^2 M_{Q}^2}
- \frac{\xi^{Q_5}_{fm} \xi^{Q_5 *}_{fm} \Yukawa{u}_{il} \Yukawadagger{u}_{kj}}{16 \pi^2 M_{Q_5}^2}
   \\
   &- \frac{\Yukawa{u}_{il} (\xi^{Q_7} \Yukawadagger{u})_{fj} \xi^{Q_7 *}_{fk}}{32 \pi^2 M_{Q_7}^2}
    - \frac{(\Yukawa{u} \xi^{Q_7 \dagger})_{if} \Yukawadagger{u})_{kj} \xi^{Q_7}_{fl}}{32 \pi^2 M_{Q_7}^2}
    - \frac{\xi^{Q_7}_{fm} \xi^{Q_7 *}_{fm} \Yukawa{u}_{il} \Yukawadagger{u}_{kj}}{16 \pi^2 M_{Q_7}^2}
    \\
   &+ \frac{\Yukawa{u}_{il} \xi^{T_1}_{fj} (\Yukawadagger{u} \xi^{T_1 \dagger})_{kf}}{128 \pi^2 M_{T_1}^2}
    + \frac{\xi^{T_1 *}_{fi} \Yukawadagger{u}_{kj} (\xi^{T_1} \Yukawa{u})_{fl}}{128 \pi^2 M_{T_1}^2}
    - \frac{\Yukawa{u}_{il} \Yukawadagger{u}_{kj} \xi^{T_1 *}_{fm} \xi^{T_1}_{fm}}{64 \pi^2 M_{T_1}^2}
    \\
   &- \frac{3 \Yukawa{u}_{il} \Yukawadagger{u}_{kj} \xi^{T_2 *}_{fm} \xi^{T_2}_{fm}}{64 \pi^2 M_{T_2}^2} 
    \,,
    \end{aligned}
\\
\left[ C_{uu} \right]_{ijkl} &= 
    \begin{aligned}[t]
   &- \frac{\xi^{u *}_{fi} \xi^{u}_{fj} \xi^{u *}_{gk} \xi^{u}_{gl}}{64 \pi^2 M_Q^2}
    + \frac{3 (\Yukawadagger{u} \Yukawa{u})_{ij} \xi^{u *}_{fk} \xi^{u}_{fl}}{128 \pi^2 M_{Q}^2}
    + \frac{3 (\Yukawadagger{u} \Yukawa{u})_{kl} \xi^{u *}_{fi} \xi^{u}_{fj}}{128 \pi^2 M_{Q}^2}
    \\
  &- \frac{\xi^{q_7 *}_{fi} \xi^{Q_7}_{fj} \xi^{Q_7 *}_{gk} \xi^{Q_7}_{gl}}{64 \pi^2 M_{Q_7}^2}
   - \frac{3 (\Yukawadagger{u} \Yukawa{u})_{ij} \xi^{Q_7 *}_{fk} \xi^{Q_7}_{fl}}{128 \pi^2 M_{Q_7}^2}
   - \frac{3 (\Yukawadagger{u} \Yukawa{u})_{kl} \xi^{Q_7 *}_{fi} \xi^{Q_7}_{fj}}{128 \pi^2 M_{Q_7}^2}
    \,,
    \end{aligned}
\end{align}
where $f,g$ are flavour indices for the new VLQs, we have assumed equal masses for multiple generations of VLQs to simplify the loop functions, and the matching conditions have been specified at the VLQ mass scale, such that $\ln (\mu^2 / M^2)$ terms vanish.

In the down-basis we have adopted, it is useful to note that products of Yukawa matrices can be simplified as
\begin{equation}
\left( Y^u Y^{u, \dagger} \right)_{ij} = y_t^2 \V*{3i} \V{3j} + \BigO{y_c^2} \,,
\quad
\left( Y^{u, \dagger} Y^u \right)_{ij} = y_t^2 \delta_{3i} \delta_{3j} + \BigO{y_c^2} \,.
\end{equation}
Note though that we keep the full expressions in our numerical analyses.

\section{Effective CKM elements}
\label{app:BSM_ckm_values}
In the presence of non-zero SMEFT coefficients, the effective CKM elements as extracted from $\beta$ decay, semi-leptonic kaon decay, and leptonic kaon and pion decay are:
\begin{align}
V_{ud}^\beta &= \V{ud} + v^2 \left[ V_\text{CKM} \cdot \CHqthree \right]_{11} + \frac{v^2}{2} \CHudbracket_{11} \,,\\
V_{us}^{K_{\ell3}} &= \V{us} + v^2 \left[ V_\text{CKM} \cdot \CHqthree \right]_{12} + \frac{v^2}{2} \CHudbracket_{12} \,,\\
V_{uq}^{M_{\mu2}} &= \V{uq} + v^2 \left[ V_\text{CKM} \cdot \CHqthree \right]_{1q} - \frac{v^2}{2} \CHudbracket_{1q}\,,
\end{align}
for $q = d,s$ and $M = K,\pi$ in the final equation.
(Notice that we obviously see here how right-handed currents are needed to resolve the tension between \V{us} determinations.)

Rearranging, and assuming small NP contributions in only the four coefficients $\CHqthreebracket_{11,12}$ and $\CHudbracket_{11,12}$ we find:
\begin{align}
\V{ud} &= V_{ud}^\beta - v^2 \left( V_{ud}^\beta \CHqthreebracket_{11} + \frac{1}{2} \CHudbracket_{11} \right) \,,\\
\V{us} &= V_{us}^{K_{\ell3}} - v^2 \left( V_{ud}^\beta \CHqthreebracket_{12} + \frac{1}{2} \CHudbracket_{12} \right) \,,\\
\frac{\V{us}}{\V{ud}} &= \frac{V_{us}^{K_{\mu2}} - v^2 \left( V_{ud} \CHqthreebracket_{12} - \frac{1}{2} \CHudbracket_{12} \right)}{V_{ud}^{\pi_{\mu2}} - v^2 \left( V_{ud} \CHqthreebracket_{11} - \frac{1}{2} \CHudbracket_{11} \right)} \\
&= \left(\frac{\V{us}}{\V{ud}}\right)^{M_{\mu2}}\!\!\!\! + v^2 \left( \left(\frac{\V{us}}{\V{ud}}\right)^{M_{\mu2}} \CHqthreebracket_{11} - \CHqthreebracket_{12} - \left(\frac{\V{us}}{\V{ud}}\right)^{M_{\mu2}} \frac{\CHudbracket_{11}}{2 V_{ud}^\beta} + \frac{\CHudbracket_{12}}{2 V_{ud}^\beta} \right)\,.
\end{align}

\section{\texorpdfstring{\smelli}{smelli} observables}
\label{app:smelli_obs}

In \cref{tab:smelli_Kl3_obs,tab:smelli_beta+n_obs,tab:smelli_tau_excl_obs,tab:smelli_Ddecay_obs,tab:smelli_K_FCNC_obs,tab:smelli_ewpo_obs} we list the observables shown in our global fits, along with the relevant experimental measurements and theory papers used in the computation.

\begin{table}
\begin{tabular}{@{}Lll|Lll@{}}
\toprule
\text{Observable} & Exp. & Theory & \text{Observable} & Exp. & Theory \\
\midrule 
\text{BR}(K_L\to \pi^+e^+\nu) & \cite{ParticleDataGroup:2022pth} & \cite{Antonelli:2010yf,Bernard:2009zm}
& \text{BR}(K_S\to \pi^+e^+\nu) & \cite{KLOE-2:2022dot} & \cite{Antonelli:2010yf,Bernard:2009zm} \\ 
\text{BR}(K^+\to \pi^0e^+\nu) & \cite{ParticleDataGroup:2022pth} & \cite{Antonelli:2010yf,Bernard:2009zm}
& \text{BR}(K_L\to \pi^+\mu^+\nu) & \cite{ParticleDataGroup:2022pth} & \cite{Antonelli:2010yf,Bernard:2009zm} \\ 
\text{BR}(K_S\to \pi^+\mu^+\nu) & \cite{KLOE-2:2019rev} & \cite{Antonelli:2010yf,Bernard:2009zm}
& \text{BR}(K^+\to \pi^0\mu^+\nu) & \cite{ParticleDataGroup:2022pth} & \cite{Antonelli:2010yf,Bernard:2009zm} \\ 
\ln(C)(K^+\to \pi^0\mu^+\nu) & \cite{Moulson:2014cra} & \cite{Antonelli:2010yf,Bernard:2009zm}
& R_T(K^+\to \pi^0\mu^+\nu) & \cite{Yushchenko:2003xz} & \cite{Antonelli:2010yf,Bernard:2009zm} \\ 
\bottomrule
\end{tabular}
\caption{The ``$K_{\ell 3}$'' observables used in \smelli, along with the relevant experimental measurements and theory papers used in the computation.}
\label{tab:smelli_Kl3_obs}
\end{table}

\begin{table}
\begin{tabular}{@{}Lll|Lll@{}}
\toprule
\text{Observable} & Exp. & Theory & \text{Observable} & Exp. & Theory \\
\midrule 
\mathcal{F}t({}^{10}\text{C}) & \cite{Hardy:2020qwl} & \cite{Gonzalez-Alonso:2018omy} 
& \mathcal{F}t({}^{14}\text{O}) & \cite{Hardy:2020qwl} & \cite{Gonzalez-Alonso:2018omy} \\ 
\mathcal{F}t({}^{22}\text{Mg}) & \cite{Hardy:2020qwl} & \cite{Gonzalez-Alonso:2018omy} 
& \mathcal{F}t({}^{26m}\text{Al}) & \cite{Hardy:2020qwl} & \cite{Gonzalez-Alonso:2018omy} \\ 
\mathcal{F}t({}^{34}\text{Cl}) & \cite{Hardy:2020qwl} & \cite{Gonzalez-Alonso:2018omy} 
& \mathcal{F}t({}^{34}\text{Ar}) & \cite{Hardy:2020qwl} & \cite{Gonzalez-Alonso:2018omy} \\ 
\mathcal{F}t({}^{38m}\text{K}) & \cite{Hardy:2020qwl} & \cite{Gonzalez-Alonso:2018omy} 
& \mathcal{F}t({}^{38}\text{Ca}) & \cite{Hardy:2020qwl} & \cite{Gonzalez-Alonso:2018omy} \\ 
\mathcal{F}t({}^{42}\text{Sc}) & \cite{Hardy:2020qwl} & \cite{Gonzalez-Alonso:2018omy} 
& \mathcal{F}t({}^{46}\text{V}) & \cite{Hardy:2020qwl} & \cite{Gonzalez-Alonso:2018omy} \\ 
\mathcal{F}t({}^{50}\text{Mn}) & \cite{Hardy:2020qwl} & \cite{Gonzalez-Alonso:2018omy} 
& \mathcal{F}t({}^{54}\text{Co}) & \cite{Hardy:2020qwl} & \cite{Gonzalez-Alonso:2018omy} \\ 
\mathcal{F}t({}^{62}\text{Ga}) & \cite{Hardy:2020qwl} & \cite{Gonzalez-Alonso:2018omy} 
& \mathcal{F}t({}^{74}\text{Rb}) & \cite{Hardy:2020qwl} & \cite{Gonzalez-Alonso:2018omy} \\ 
\tau_n & \cite{UCNt:2021pcg} & \cite{Gonzalez-Alonso:2018omy} 
& \tilde{A}_n & \cite{Markisch:2018ndu,Brown:2017mhw,Mund:2012fq} & \cite{Gonzalez-Alonso:2018omy} \\ 
R_n & \cite{Kozela:2011mc} & \cite{Gonzalez-Alonso:2018omy} 
& \lambda_{AB} & \cite{Mostovoi:2001ye} & \cite{Gonzalez-Alonso:2018omy} \\ 
a_n & \cite{Gonzalez-Alonso:2018omy} & \cite{Gonzalez-Alonso:2018omy} 
& \tilde{a}_n & \cite{Darius:2017arh} & \cite{Gonzalez-Alonso:2018omy} \\ 
\tilde{B}_n & \cite{Gonzalez-Alonso:2018omy} & \cite{Gonzalez-Alonso:2018omy} 
& D_n & \cite{Gonzalez-Alonso:2018omy} & \cite{Gonzalez-Alonso:2018omy} \\ 
\bottomrule
\end{tabular}
\caption{The ``beta'' observables used in \smelli, along with the relevant experimental measurements and theory papers used in the computation.}
\label{tab:smelli_beta+n_obs}
\end{table}

\begin{table}
\begin{tabular}{@{}Ll|Ll@{}}
\toprule
\text{Observable} & Exp. &  \text{Observable} & Exp.  \\
\midrule 
\text{BR}(\tau^+\to \pi^+\bar\nu) & \cite{ParticleDataGroup:2022pth} & \text{BR}(\tau^+\to K^+\bar\nu) & \cite{ParticleDataGroup:2022pth}  \\ 
\bottomrule
\end{tabular}
\caption{The exclusive $\tau$ decays observables used in \smelli, along with the relevant experimental measurements and theory papers used in the computation.}
\label{tab:smelli_tau_excl_obs}
\end{table}

\begin{table}
\begin{tabular}{@{}Lll|Lll@{}}
\toprule
\text{Observable} & Exp. & Theory & \text{Observable} & Exp. & Theory \\
\midrule 
\text{BR}(D^+\to e^+\nu_e) & \cite{ParticleDataGroup:2022pth} &  
&\text{BR}(D^+\to \mu^+\nu_\mu) & \cite{ParticleDataGroup:2022pth} & 
\\ 
\text{BR}(D^+\to \tau^+\nu_\tau) & \cite{ParticleDataGroup:2022pth} &  
&\text{BR}(D_s\to e^+\nu_e) & \cite{ParticleDataGroup:2022pth} & 
\\ 
\text{BR}(D_s\to \mu^+\nu_\mu) & \cite{ParticleDataGroup:2022pth} &  
&\text{BR}(D_s\to \tau^+\nu_\tau) & \cite{ParticleDataGroup:2022pth} & 
\\ 
\text{BR}(D^0\to \pi^- \mu^+\nu_\mu) & \cite{ParticleDataGroup:2022pth} & \cite{Gubernari:2018wyi} 
&\text{BR}(D^0\to \pi^- e^+\nu_e) & \cite{ParticleDataGroup:2022pth} & \cite{Gubernari:2018wyi} 
\\ 
\text{BR}(D^0\to K^- e^+\nu_e) & \cite{ParticleDataGroup:2022pth} & \cite{Gubernari:2018wyi} 
&\text{BR}(D^0\to K^- \mu^+\nu_\mu) & \cite{ParticleDataGroup:2022pth} & \cite{Gubernari:2018wyi} 
\\
\langle\text{BR}\rangle(D^0\to \pi^- e^+\nu_e) & \cite{CLEO:2009svp,BESIII:2015tql} & \cite{Gubernari:2018wyi} 
&\text{BR}(D^+\to \pi^0\mu^+\nu_\mu) & \cite{ParticleDataGroup:2022pth} & \cite{Gubernari:2018wyi}
\\ 
\langle\text{BR}\rangle(D^0\to K^- e^+\nu_e) & \cite{CLEO:2009svp,BESIII:2015tql} & \cite{Gubernari:2018wyi}
&\text{BR}(D^+\to K^0\mu^+\nu_\mu) & \cite{ParticleDataGroup:2022pth} & \cite{Gubernari:2018wyi} 
\\ 
\langle\text{BR}\rangle(D^+\to K^0e^+\nu_e) & \cite{CLEO:2009svp,BESIII:2017ylw} & \cite{Gubernari:2018wyi} 
&\text{BR}(D^+\to \pi^0e^+\nu_e) & \cite{ParticleDataGroup:2022pth} & \cite{Gubernari:2018wyi} 
\\ 
\langle\text{BR}\rangle(D^+\to \pi^0e^+\nu_e) & \cite{CLEO:2009svp,BESIII:2017ylw} & \cite{Gubernari:2018wyi} 
&\text{BR}(D^+\to K^0e^+\nu_e) & \cite{ParticleDataGroup:2022pth} & \cite{Gubernari:2018wyi} 
\\ 
\bottomrule
\end{tabular}
\caption{The $D$ decay observables used in \smelli, along with the relevant experimental measurements and theory papers used in the computation.
$\langle \text{BR} \rangle$ are $q^2$-binned branching ratios.}
\label{tab:smelli_Ddecay_obs}
\end{table}

\begin{table}
\begin{tabular}{@{}Lll|Lll@{}}
\toprule
\text{Observable} & Exp. & Theory & \text{Observable} & Exp. & Theory \\
\midrule 
\text{BR}(K^+\to\pi^+\nu\bar\nu) & \cite{NA62:2021zjw} & \cite{Brod:2010hi,Buras:2006gb,Buchalla:1998ba,Isidori:2005xm,Mescia:2007kn} & \text{BR}(K_L\to\pi^0\nu\bar\nu) & \cite{Ahn:2018mvc} & \cite{Brod:2010hi,Buras:2006gb,Buchalla:1998ba,Isidori:2005xm,Mescia:2007kn} \\ 
\text{BR}(K_L\to e^+e^-) & \cite{ParticleDataGroup:2022pth} & \cite{Bobeth:2013uxa,Gorbahn:2006bm,Chobanova:2017rkj} & \text{BR}(K_S\to e^+e^-) & \cite{ParticleDataGroup:2022pth} & \cite{Bobeth:2013uxa,Gorbahn:2006bm,Chobanova:2017rkj} \\ 
\text{BR}(K_L\to \mu^+\mu^-) & \cite{ParticleDataGroup:2022pth} & \cite{Bobeth:2013uxa,Gorbahn:2006bm,Chobanova:2017rkj} & \text{BR}(K_S\to \mu^+\mu^-) & \cite{LHCb:2020ycd} & \cite{Bobeth:2013uxa,Gorbahn:2006bm,Chobanova:2017rkj} \\ 
\vert\epsilon_K\vert & \cite{ParticleDataGroup:2022pth} & \cite{Blum:2011ng,Brod:2011ty,Brod:2010mj,Carrasco:2015pra} \\ 
\bottomrule
\end{tabular}
\caption{The ``$K$ FCNC'' observables used in \smelli, along with the relevant experimental measurements and theory papers used in the computation.}
\label{tab:smelli_K_FCNC_obs}
\end{table}

\begin{table}
\begin{tabular}{@{}Lll|Lll@{}}
\toprule
\text{Observable} & Exp. & Theory & \text{Observable} & Exp. & Theory \\
\midrule 
\Gamma_Z & \cite{Janot:2019oyi} & \cite{Brivio:2017vri,Freitas:2014hra} 
& \sigma_\text{had}^0 & \cite{Janot:2019oyi} & \cite{Brivio:2017vri,Freitas:2014hra} \\ 
R_ e^0 & \cite{Janot:2019oyi} & \cite{Brivio:2017vri,Freitas:2014hra} 
& R_\mu^0 & \cite{Janot:2019oyi} & \cite{Brivio:2017vri,Freitas:2014hra} \\ 
R_\tau^0 & \cite{Janot:2019oyi} & \cite{Brivio:2017vri,Freitas:2014hra} 
& A_\text{FB}^{0, e} & \cite{Janot:2019oyi} & \cite{Brivio:2017vri} \\ 
A_\text{FB}^{0,\mu} & \cite{Janot:2019oyi} & \cite{Brivio:2017vri} 
& A_\text{FB}^{0,\tau} & \cite{Janot:2019oyi} & \cite{Brivio:2017vri} \\ 
A_ e & \cite{ALEPH:2005ab} & \cite{Brivio:2017vri} 
& A_\mu & \cite{ALEPH:2005ab} & \cite{Brivio:2017vri} \\ 
A_\tau & \cite{ALEPH:2005ab} & \cite{Brivio:2017vri} 
& R_ b^0 & \cite{ALEPH:2005ab} & \cite{Brivio:2017vri,Freitas:2014hra} \\ 
R_ c^0 & \cite{ALEPH:2005ab} & \cite{Brivio:2017vri,Freitas:2014hra} 
& A_\text{FB}^{0, b} & \cite{ALEPH:2005ab} & \cite{Brivio:2017vri} \\ 
A_\text{FB}^{0, c} & \cite{ALEPH:2005ab} & \cite{Brivio:2017vri} 
& A_ b & \cite{ALEPH:2005ab} & \cite{Brivio:2017vri} \\ 
A_ c & \cite{ALEPH:2005ab} & \cite{Brivio:2017vri} 
& m_W & \cite{Aaltonen:2013iut,Aaboud:2017svj} & \cite{Awramik:2003rn,Bjorn:2016zlr,Brivio:2017vri} \\ 
\Gamma_W & \cite{ParticleDataGroup:2022pth} & \cite{Brivio:2017vri} 
& \text{BR}(W^\pm\to  e^\pm\nu) & \cite{Schael:2013ita} & \cite{Brivio:2017vri} \\ 
\text{BR}(W^\pm\to \mu^\pm\nu) & \cite{Schael:2013ita} & \cite{Brivio:2017vri} 
& \text{BR}(W^\pm\to \tau^\pm\nu) & \cite{Schael:2013ita} & \cite{Brivio:2017vri} \\ 
\text{R}(W^+\to cX) & \cite{ParticleDataGroup:2022pth} & \cite{Brivio:2017vri} 
& \text{R}_{\mu  e}(W^\pm\to \ell^\pm\nu) & \cite{Aaij:2016qqz} & \cite{Brivio:2017vri} \\ 
\text{R}_{\tau  e}(W^\pm\to \ell^\pm\nu) & \cite{Abbott:1999pk} & \cite{Brivio:2017vri} 
& \text{R}_{\tau \mu}(W^\pm\to \ell^\pm\nu) & \cite{Aad:2020ayz} & \cite{Brivio:2017vri} \\ 
A_ s & \cite{Abe:2000uc} & \cite{Brivio:2017vri} 
& R_{uc}^0 & \cite{ParticleDataGroup:2022pth} & \cite{Brivio:2017vri,Freitas:2014hra} \\ 
\bottomrule
\end{tabular}
\caption{The EWPO observables used in \smelli, along with the relevant experimental measurements and theory papers used in the computation.}
\label{tab:smelli_ewpo_obs}
\end{table}


\section{Plots for 4-D EFT scenarios}
\label{app:4d_scenarios}

\begin{figure}
\includegraphics[width=0.7\textwidth]{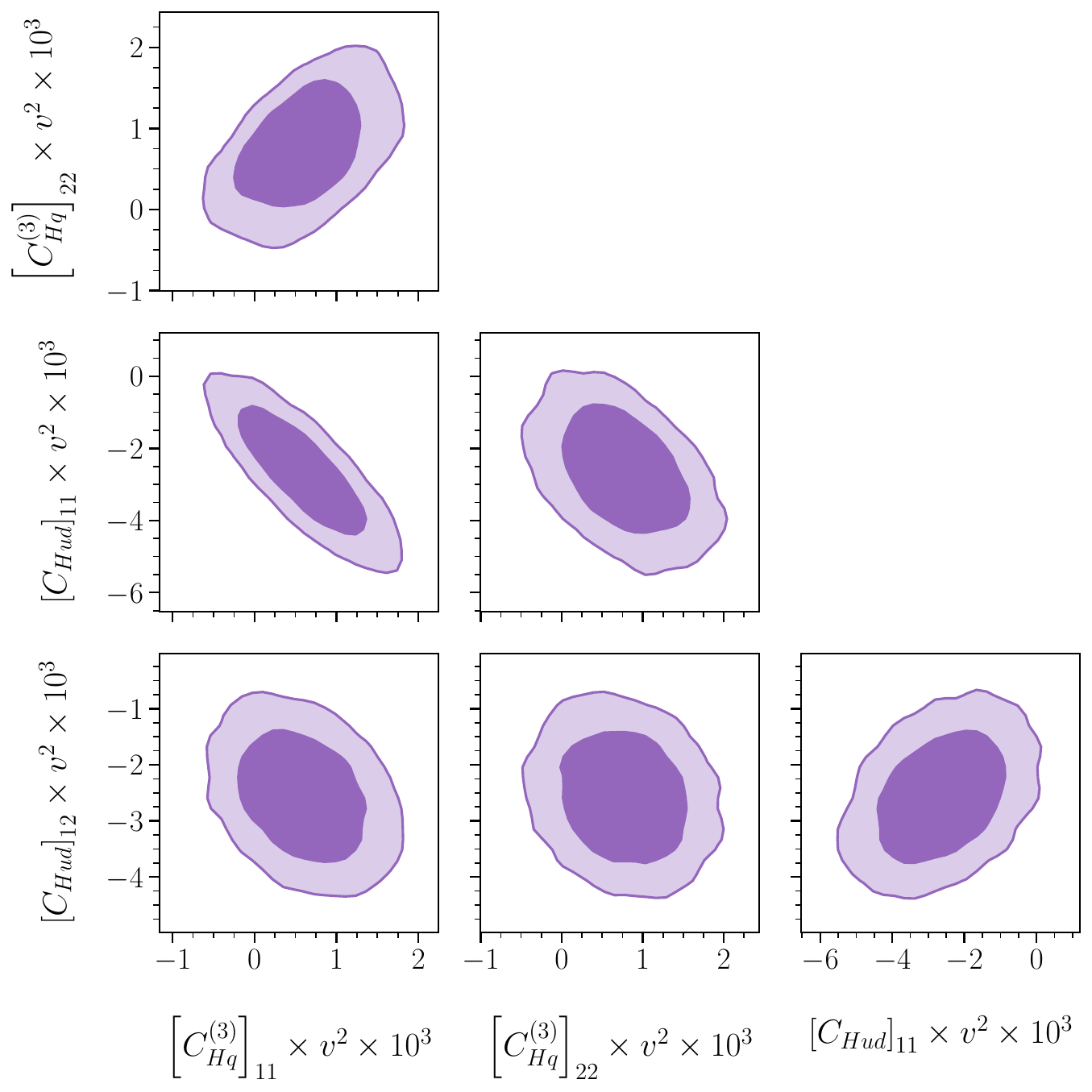}
\caption{4-D scenario with  $\CHqthreebracket_{11}$, $\CHqthreebracket_{22}$, $\CHudbracket_{11}$, and $\CHudbracket_{12}$ being non-zero.}
\label{fig:eft_LH_RH_flavour_diagonal}
\end{figure}

\begin{figure}
\includegraphics[width=0.7\textwidth]{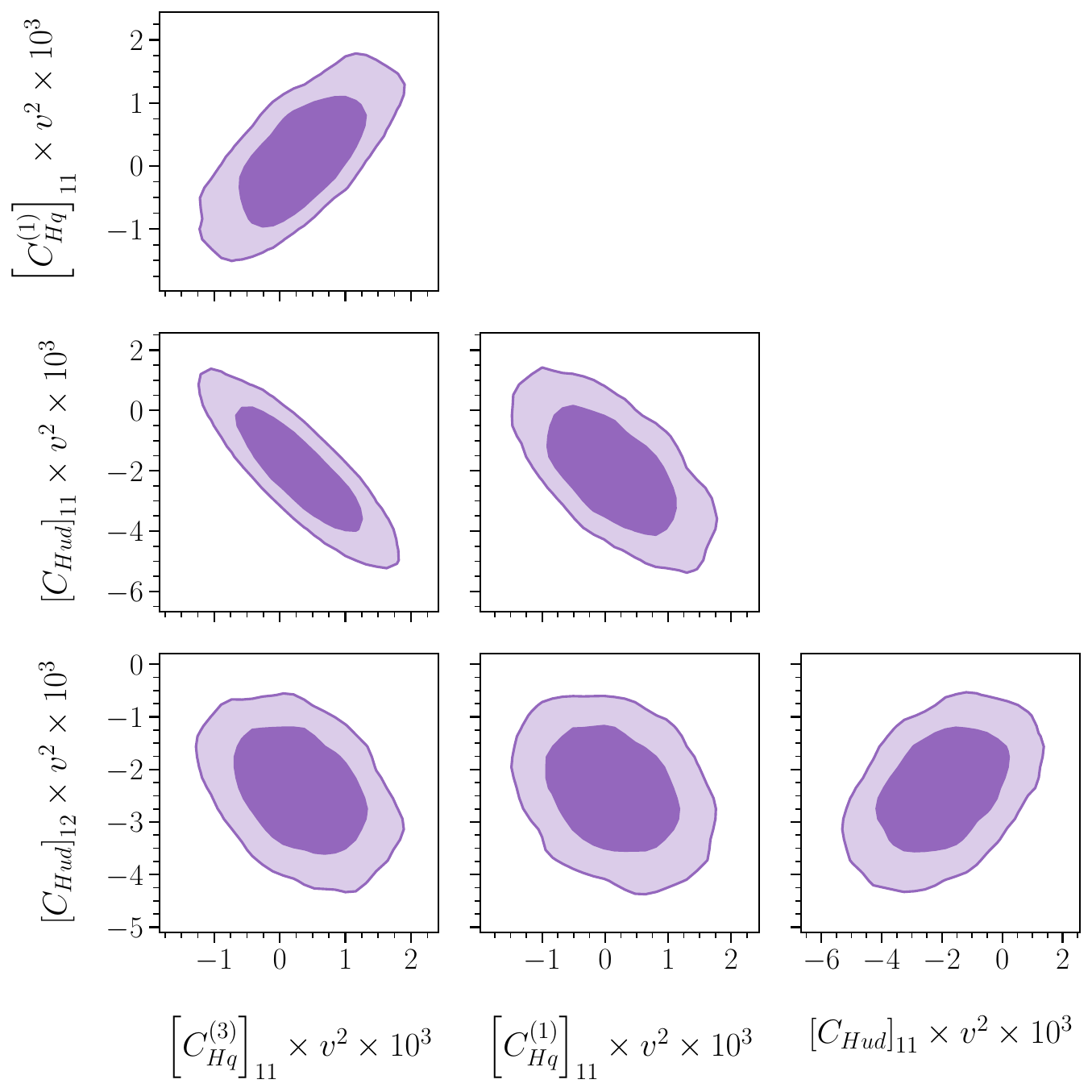}
\caption{4-D scenario with $\CHqthreebracket_{11}$, $\CHqonebracket_{11}$, $\CHudbracket_{11}$, and $\CHudbracket_{12}$ being non-zero.}
\label{fig:eft_LH_RH_singlet_triplet}
\end{figure}

As \Dz--\Dzb mixing is a serious bound when considering left-handed modifications, we go beyond the 3-D scenario by allowing a $U(2)$ flavour symmetry for $\CHqthree$ as well as the possibility of cancellation between $\CHqthree$ and $\CHqone$ in $Z$ couplings to up quarks, as done earlier in our 1-D scenarios.
We therefore performed a global fit to the two scenarios
\begin{enumerate}
    \item $\CHqthreebracket_{11}$,  $\CHudbracket_{11}$, $\CHudbracket_{12}$ plus $\CHqthreebracket_{22}$, with the results shown in \cref{fig:eft_LH_RH_flavour_diagonal}
    \item[] and 
    \item $\CHqthreebracket_{11}$,  $\CHudbracket_{11}$, $\CHudbracket_{12}$ plus $\CHqonebracket_{11}$, with the results shown in \cref{fig:eft_LH_RH_singlet_triplet}.
\end{enumerate}

\footnotesize

\bibliographystyle{utphys28mod} 
\bibliography{BIB}

\end{document}